\documentclass[submission,copyright,creativecommons]{eptcs}
\providecommand{\event}{MSFP 2020} 
\usepackage{breakurl}              
\usepackage{underscore}            

\usepackage{amsfonts}
\usepackage{amsmath}
\usepackage{amsthm}
\usepackage{todonotes}

\usepackage[bb=boondox]{mathalfa}

\usepackage{stmaryrd}

\usepackage[all,british]{foreign}

\usepackage{microtype}

\usepackage{agda}

\usepackage[utf8]{inputenc}
\usepackage[T1]{fontenc}

\usepackage{ifthen}

\newcommand{\Nat}{\ensuremath{\mathbb{N}}}

\newcommand{\prodt}[2]{\ensuremath{\prod\nolimits_{\textstyle #1}{#2}}}
\newcommand{\sumt}[2]{\ensuremath{\sum\nolimits_{\textstyle #1}{#2}}}
\newcommand{\contint}[1]{\ensuremath{\left\llbracket #1 \right\rrbracket_{\lhd}}}

\usepackage{newunicodechar}
\DeclareUnicodeCharacter{2115}{\ensuremath{\mathbb N}}
\DeclareUnicodeCharacter{2081}{\ensuremath{_1}}
\newunicodechar{₂}{\ensuremath{\mathnormal{_2}}}
\newunicodechar{₃}{\ensuremath{\mathnormal{_3}}}
\newunicodechar{ₙ}{\ensuremath{\mathnormal{_n}}}
\newunicodechar{≥}{\ensuremath{\mathnormal\ge}}
\newunicodechar{λ}{\ensuremath{\mathnormal\lambda}}
\newunicodechar{←}{\ensuremath{\mathnormal\from}}
\newunicodechar{→}{\ensuremath{\mathnormal\to}}
\newunicodechar{∀}{\ensuremath{\mathnormal\forall}}
\newunicodechar{⟦}{\ensuremath{\mathnormal{\llbracket}}}
\newunicodechar{⟧}{\ensuremath{\mathnormal{\rrbracket}}}
\newunicodechar{Σ}{\ensuremath{\mathnormal\Sigma}}
\newunicodechar{⋄}{\ensuremath{\mathnormal{\diamond}}}
\newunicodechar{γ}{\ensuremath{\mathnormal{\gamma}}}
\newunicodechar{⊤}{\ensuremath{\mathnormal{\top}}}
\newunicodechar{◃}{\ensuremath{\mathnormal{\lhd}}}
\newunicodechar{∷}{\ensuremath{\mathnormal{::}}}
\newunicodechar{⟨}{\ensuremath{\mathnormal{\langle}}}
\newunicodechar{⟩}{\ensuremath{\mathnormal{\rangle}}}
\newunicodechar{≡}{\ensuremath{\mathnormal{\equiv}}}
\newunicodechar{×}{\ensuremath{\mathnormal{\times}}}
\newunicodechar{∘}{\ensuremath{\mathnormal{\circ}}}

\title{Multi-dimensional Arrays with Levels}
\author{Artjoms {\v{S}}inkarovs
\institute{School of Mathematical and Computer Sciences\\
Heriot-Watt University\\
Scotland}
\email{tema@pm.me}
}
\def\titlerunning{Multi-dimensional Arrays with Levels}
\def\authorrunning{A. {\v{S}}inkarovs}

\begin{document}
\maketitle

\begin{abstract}
We explore a data structure that generalises rectangular multi-dimensional
arrays.  The shape of an $n$-dimensional array is typically given by a tuple
of $n$ natural numbers.  Each element in that tuple defines the length of the
corresponding axis.  If we treat this tuple as an array, the shape of that
array is described by the single natural number $n$.  A natural number itself can
be also treated as an array with the shape described by the natural
number $1$ (or the element of any singleton set).  
This observation gives rise to the hierarchy of array types where the shape
of an array of level $l+1$ is a level-$l$ array of natural numbers.  Such a
hierarchy occurs naturally when treating arrays as containers, which 
makes it possible to define both rank- and level-polymorphic operations.  The
former can be found in most 
array languages, whereas the latter
gives rise to partial selections on a large set of hyperplanes,
which is often useful in practice.  In this 
paper we present 
an Agda formalisation of arrays with levels.  We show that the proposed
formalism supports standard rank-polymorphic array operations, while 
type system gives static guarantees that indexing is within bounds. 
We generalise the notion of ranked operator so that it becomes applicable
on arrays of arbitrary levels and we show why this may be useful in practice.
\end{abstract}

\section{Introduction}

A large number of high-performance numerical problems use
multi-dimensional arrays (often referred as tensors) as a key data structure.
On the one hand, the multi-dimensional array is a natural abstraction of a space with a regular
structure; on the other hand, computations on arrays can be efficiently
implemented on conventional computing architectures.

In functional programming, arrays typically do not get a lot of
attention, as most of computations on arrays can be expressed
as computations on nested lists or vectors, both of which are simpler
data structures.  While the latter is true, nested vectors miss a very
essential feature of many array languages --- rank polymorphism.  This is
the ability to define operations on arrays of arbitrarily many dimensions.

The tradition of rank-polymorphic programming starts from APL~\cite{Iverson:1962:APL}
and is picked-up by a number of descendants such as J~\cite{jlang}, K~\cite{klang},
FISH~\cite{fish} and others.  Rank-polymorphic array programming has
found its way into functional langauges as well.  For example, SaC~\cite{sac},
Remora~\cite{remora}, Qube~\cite{qube} are all array-based languages supporting
rank-polymorphism.  However, enforcing static safety guarantees such as lack of out-of-bound
indexing turns out to be a very challenging problem.  None of the 
functional languages above are capable of enforcing such guarantees for the full range of APL
operators.   The reason for this is that a number of these operators introduce
a dependency between the value of the input and the shape of the output.
For example, the \emph{take} operator selects a subarray from the given array,
and the shape of the subarray comes as an argument.  Therefore, a rank-polymorphic
language of APL's expressive power which guarantees correct indexing
has to support dependent types.  Most of practical languages find
such a constraint too harsh, due mainly to the fact that one has to
give-up global type inference.  Consequently, these languages make 
compromises, either with the range of supported primitives, or with type safety.

When designing a type system for a rank-polymorphic array language
the notion of containers~\cite{containers} comes in very handy.  An arrays can be thought
of as a tabulated index-value functions, where the set of valid indices
into the array is defined by the array shape --- exactly the abstraction
that containers are designed to handle.

While formalising rectangular multi-dimensional arrays using containers
(see Section~\ref{sec:cont}), we discovered a new container operation that
gives rise to the desired structure:

\begin{equation}
    (A \lhd B) \mathop{\diamond} (C \lhd D)
    = \contint{A \lhd B}C\ \mathop{\lhd}\  \lambda (a, s) \to \prodt{B\ a}{D \circ s}
\end{equation}
Using this operation, we can define a multi-dimensional array with elements
of type $X$ as:
\begin{equation}
    Array\ X =  \contint{(\Nat \lhd Fin) \diamond (\Nat \lhd Fin)}\ X
\end{equation}
We explain the derivation of the operation, and the array structure
in Section~\ref{sec:arr-as-cont}.

As $\diamond$ is a general container operation, we notice
that $- \diamond (\Nat \lhd Fin)$ can be iterated.  By doing so, we get a
hierarchy of array types that to our knowledge have not been studied before.
Intuitively they can be described as follows.  The container $\Nat \lhd Fin$ 
describes a finite 
vector, the shape of which is given
by a natural number $n$.  The indices into this vector are natural numbers that
are less than $n$. 
We call these objects level-1 arrays.  If we apply
$- \diamond (\Nat \lhd Fin)$ to a level-1 array 
we get level-2 arrays.  The shape of such a thing is a vector of natural numbers,
and the indices are vectors of natural numbers of the same size as the shape,
where each element is less than the corresponding element in the shape vector.
At the next application, level-3 arrays have shapes that are described by level-2
arrays of natural numbers, and level-3 indices are level-2 arrays of natural
numbers, where each element is less than the corresponding element in the shape.
And so on.

Beyond simple curiosity, it turns out that these higher-level arrays can
be useful in practice.  To understand why this is the case, consider
the following intuition.  The main advantage of multi-dimensional arrays
is the availability of proximity metrics for a given element.  Within
a vector, we can 
refer only to the left and right neighbours of a cell, whereas within
an $n$-dimensional array we can
refer to $2n$ neighbours.  Within a multi-level arrays, it is possible
to do exactly
the same at the level of shapes.  The shapes of level-2 arrays are always vectors,
therefore one can only talk about left and right shape neighbours.  With level-$n$
arrays we can talk about $2n$ shape neighbours, applying all the arsenal of
multi-dimensional array operations at the level of shapes.  This additional
information in the shape makes it possible to define a new class of generic
array operations that reshuffle array elements or perform non-trivial partial
selections, both of which lie at the very core of array programming.

The paper is a literate Agda script. The contributions are as follows:
\begin{itemize}
        \item Description of novel data structures that generalise multi-dimensional
                arrays;
        \item Formal definition of the data structure and the standard
                array operations in Agda (available at~\cite{github});
        \item Generalisation of rank-polymorphic array operations to
                level-polymorphism, and demonstration of the benefits in practice.
\end{itemize}

\section{Arrays as Containers\label{sec:arr-as-cont}}

In this section we briefly introduce container types and explain how
we derived the previously mentioned $\diamond$ operation.

\begin{code}[hide]%
\>[0]\AgdaKeyword{open}\AgdaSpace{}%
\AgdaKeyword{import}\AgdaSpace{}%
\AgdaModule{Data.Product}\<%
\\
\>[0]\AgdaKeyword{open}\AgdaSpace{}%
\AgdaKeyword{import}\AgdaSpace{}%
\AgdaModule{Data.Nat}\<%
\\
\>[0]\AgdaKeyword{open}\AgdaSpace{}%
\AgdaKeyword{import}\AgdaSpace{}%
\AgdaModule{Data.Unit}\<%
\\
\>[0]\AgdaKeyword{open}\AgdaSpace{}%
\AgdaKeyword{import}\AgdaSpace{}%
\AgdaModule{Data.Empty}\<%
\\
\>[0]\AgdaKeyword{open}\AgdaSpace{}%
\AgdaKeyword{import}\AgdaSpace{}%
\AgdaModule{Data.Fin}\<%
\\
\>[0]\AgdaKeyword{open}\AgdaSpace{}%
\AgdaKeyword{import}\AgdaSpace{}%
\AgdaModule{Relation.Binary.PropositionalEquality}\<%
\\
\>[0]\AgdaKeyword{open}\AgdaSpace{}%
\AgdaKeyword{import}\AgdaSpace{}%
\AgdaModule{Data.Sum}\<%
\end{code}

\subsection{Containers\label{sec:cont}}

Containers can be seen as a conceptual tool 
to describe 
``collections
of things'' such as lists or trees in a uniform way. Mathematically, containers are
endofunctors on a category of types, that
are coproducts of type-indexed families of representable functors.

We define containers
by a type of shapes $Sh$ and an $Sh$-indexed type family $Po$.
The interpretation (sometimes called extension) of a container type
is a dependent pair type where the first element is the shape of type $Sh$,
and the second element is a function from positions of that shape to
the element type.  Following Conor McBride's syntax, we specify containers
in Agda as follows\footnote{For presentational purposes we avoid level
polymorphism and force shapes to be elements of \texttt{Set}.  A level-polymorphic
version of containers can be found in Agda's standard library.}:

\begin{code}%
\>[0]\AgdaKeyword{record}\AgdaSpace{}%
\AgdaRecord{Con}\AgdaSpace{}%
\AgdaSymbol{:}\AgdaSpace{}%
\AgdaPrimitiveType{Set₁}\AgdaSpace{}%
\AgdaKeyword{where}\<%
\\
\>[0][@{}l@{\AgdaIndent{0}}]%
\>[2]\AgdaKeyword{constructor}\AgdaSpace{}%
\AgdaOperator{\AgdaInductiveConstructor{\AgdaUnderscore{}◃\AgdaUnderscore{}}}\<%
\\
\>[2]\AgdaKeyword{field}\<%
\\
\>[2][@{}l@{\AgdaIndent{0}}]%
\>[4]\AgdaField{Sh}\AgdaSpace{}%
\AgdaSymbol{:}\AgdaSpace{}%
\AgdaPrimitiveType{Set}\<%
\\
\>[4]\AgdaField{Po}\AgdaSpace{}%
\AgdaSymbol{:}\AgdaSpace{}%
\AgdaBound{Sh}\AgdaSpace{}%
\AgdaSymbol{→}\AgdaSpace{}%
\AgdaPrimitiveType{Set}\<%
\\
\>[2]\AgdaOperator{\AgdaFunction{⟦\AgdaUnderscore{}$\rrbracket_{\lhd}$}}%
\>[8]\AgdaSymbol{:}\AgdaSpace{}%
\AgdaPrimitiveType{Set}\AgdaSpace{}%
\AgdaSymbol{→}\AgdaSpace{}%
\AgdaPrimitiveType{Set}\<%
\\
\>[2]\AgdaOperator{\AgdaFunction{⟦\AgdaUnderscore{}$\rrbracket_{\lhd}$}}\AgdaSpace{}%
\AgdaBound{X}\AgdaSpace{}%
\AgdaSymbol{=}\AgdaSpace{}%
\AgdaRecord{Σ}\AgdaSpace{}%
\AgdaField{Sh}\AgdaSpace{}%
\AgdaSymbol{λ}\AgdaSpace{}%
\AgdaBound{s}\AgdaSpace{}%
\AgdaSymbol{→}\AgdaSpace{}%
\AgdaField{Po}\AgdaSpace{}%
\AgdaBound{s}\AgdaSpace{}%
\AgdaSymbol{→}\AgdaSpace{}%
\AgdaBound{X}\<%
\end{code}
To develop an intuition, consider lists of $X$s, expressed as a container.
\begin{code}[hide]%
\>[0]\AgdaKeyword{open}\AgdaSpace{}%
\AgdaModule{Con}\AgdaSpace{}%
\AgdaKeyword{public}\<%
\\
\>[0]\AgdaFunction{List}\AgdaSpace{}%
\AgdaSymbol{:}\AgdaSpace{}%
\AgdaPrimitiveType{Set}\AgdaSpace{}%
\AgdaSymbol{→}\AgdaSpace{}%
\AgdaPrimitiveType{Set}\<%
\\
\>[0]\AgdaFunction{Tree}\AgdaSpace{}%
\AgdaSymbol{:}\AgdaSpace{}%
\AgdaPrimitiveType{Set}\AgdaSpace{}%
\AgdaSymbol{→}\AgdaSpace{}%
\AgdaPrimitiveType{Set}\<%
\end{code}
\begin{code}%
\>[0]\AgdaFunction{List}\AgdaSpace{}%
\AgdaBound{X}\AgdaSpace{}%
\AgdaSymbol{=}\AgdaSpace{}%
\AgdaOperator{\AgdaFunction{⟦}}\AgdaSpace{}%
\AgdaDatatype{ℕ}\AgdaSpace{}%
\AgdaOperator{\AgdaInductiveConstructor{◃}}\AgdaSpace{}%
\AgdaDatatype{Fin}\AgdaSpace{}%
\AgdaOperator{\AgdaFunction{$\rrbracket_{\lhd}$}}\AgdaSpace{}%
\AgdaBound{X}\AgdaSpace{}%
\AgdaComment{\ {-}{-} ≡ Σ ℕ λ n → Fin n → X}\<%
\end{code}
For any given length $n$, the data of the list is modeled
by a function of type \texttt{Fin n → X}.  While the type does not carry the length as an
argument, each item of the list container type carries the length in the first
element of the dependent pair.

Consider binary trees defined as containers.
\begin{code}%
\>[0]\AgdaKeyword{data}\AgdaSpace{}%
\AgdaDatatype{Tr}\AgdaSpace{}%
\AgdaSymbol{:}\AgdaSpace{}%
\AgdaPrimitiveType{Set}\AgdaSpace{}%
\AgdaKeyword{where}\<%
\\
\>[0][@{}l@{\AgdaIndent{0}}]%
\>[2]\AgdaInductiveConstructor{Empty}\AgdaSpace{}%
\AgdaSymbol{:}\AgdaSpace{}%
\AgdaDatatype{Tr}\<%
\\
\>[2]\AgdaInductiveConstructor{Node}\AgdaSpace{}%
\AgdaSymbol{:}\AgdaSpace{}%
\AgdaDatatype{Tr}\AgdaSpace{}%
\AgdaSymbol{→}\AgdaSpace{}%
\AgdaDatatype{Tr}\AgdaSpace{}%
\AgdaSymbol{→}\AgdaSpace{}%
\AgdaDatatype{Tr}\<%
\\
\\[\AgdaEmptyExtraSkip]%
\>[0]\AgdaKeyword{data}\AgdaSpace{}%
\AgdaDatatype{Tx}\AgdaSpace{}%
\AgdaSymbol{:}\AgdaSpace{}%
\AgdaDatatype{Tr}\AgdaSpace{}%
\AgdaSymbol{→}\AgdaSpace{}%
\AgdaPrimitiveType{Set}\AgdaSpace{}%
\AgdaKeyword{where}\<%
\\
\>[0][@{}l@{\AgdaIndent{0}}]%
\>[2]\AgdaInductiveConstructor{Done}\AgdaSpace{}%
\AgdaSymbol{:}\AgdaSpace{}%
\AgdaSymbol{∀}\AgdaSpace{}%
\AgdaSymbol{\{}\AgdaBound{l}\AgdaSpace{}%
\AgdaBound{r}\AgdaSymbol{\}}\AgdaSpace{}%
\AgdaSymbol{→}\AgdaSpace{}%
\AgdaDatatype{Tx}\AgdaSpace{}%
\AgdaSymbol{(}\AgdaInductiveConstructor{Node}\AgdaSpace{}%
\AgdaBound{l}\AgdaSpace{}%
\AgdaBound{r}\AgdaSymbol{)}\<%
\\
\>[2]\AgdaOperator{\AgdaInductiveConstructor{R\AgdaUnderscore{}}}%
\>[7]\AgdaSymbol{:}\AgdaSpace{}%
\AgdaSymbol{∀}\AgdaSpace{}%
\AgdaSymbol{\{}\AgdaBound{l}\AgdaSpace{}%
\AgdaBound{r}\AgdaSymbol{\}}\AgdaSpace{}%
\AgdaSymbol{→}\AgdaSpace{}%
\AgdaDatatype{Tx}\AgdaSpace{}%
\AgdaBound{r}\AgdaSpace{}%
\AgdaSymbol{→}\AgdaSpace{}%
\AgdaDatatype{Tx}\AgdaSpace{}%
\AgdaSymbol{(}\AgdaInductiveConstructor{Node}\AgdaSpace{}%
\AgdaBound{l}\AgdaSpace{}%
\AgdaBound{r}\AgdaSymbol{)}\<%
\\
\>[2]\AgdaOperator{\AgdaInductiveConstructor{L\AgdaUnderscore{}}}%
\>[7]\AgdaSymbol{:}\AgdaSpace{}%
\AgdaSymbol{∀}\AgdaSpace{}%
\AgdaSymbol{\{}\AgdaBound{l}\AgdaSpace{}%
\AgdaBound{r}\AgdaSymbol{\}}\AgdaSpace{}%
\AgdaSymbol{→}\AgdaSpace{}%
\AgdaDatatype{Tx}\AgdaSpace{}%
\AgdaBound{l}\AgdaSpace{}%
\AgdaSymbol{→}\AgdaSpace{}%
\AgdaDatatype{Tx}\AgdaSpace{}%
\AgdaSymbol{(}\AgdaInductiveConstructor{Node}\AgdaSpace{}%
\AgdaBound{l}\AgdaSpace{}%
\AgdaBound{r}\AgdaSymbol{)}\<%
\\
\\[\AgdaEmptyExtraSkip]%
\>[0]\AgdaFunction{Tree}\AgdaSpace{}%
\AgdaBound{X}\AgdaSpace{}%
\AgdaSymbol{=}\AgdaSpace{}%
\AgdaOperator{\AgdaFunction{⟦}}\AgdaSpace{}%
\AgdaDatatype{Tr}\AgdaSpace{}%
\AgdaOperator{\AgdaInductiveConstructor{◃}}\AgdaSpace{}%
\AgdaDatatype{Tx}\AgdaSpace{}%
\AgdaOperator{\AgdaFunction{$\rrbracket_{\lhd}$}}\AgdaSpace{}%
\AgdaBound{X}\<%
\end{code}
\begin{code}[hide]%
\>[0]\AgdaComment{\ {-}{-} Here we explore the equality from the paper.}\<%
\\
\>[0]\AgdaOperator{\AgdaFunction{\AgdaUnderscore{}⋄\AgdaUnderscore{}}}\AgdaSpace{}%
\AgdaSymbol{:}\AgdaSpace{}%
\AgdaRecord{Con}\AgdaSpace{}%
\AgdaSymbol{→}\AgdaSpace{}%
\AgdaRecord{Con}\AgdaSpace{}%
\AgdaSymbol{→}\AgdaSpace{}%
\AgdaRecord{Con}\<%
\\
\>[0]\AgdaSymbol{(}\AgdaBound{S}\AgdaSpace{}%
\AgdaOperator{\AgdaInductiveConstructor{◃}}\AgdaSpace{}%
\AgdaBound{P}\AgdaSymbol{)}\AgdaSpace{}%
\AgdaOperator{\AgdaFunction{⋄}}\AgdaSpace{}%
\AgdaSymbol{(}\AgdaBound{S₁}\AgdaSpace{}%
\AgdaOperator{\AgdaInductiveConstructor{◃}}\AgdaSpace{}%
\AgdaBound{P₁}\AgdaSymbol{)}\AgdaSpace{}%
\AgdaSymbol{=}\AgdaSpace{}%
\AgdaOperator{\AgdaFunction{⟦}}\AgdaSpace{}%
\AgdaBound{S}\AgdaSpace{}%
\AgdaOperator{\AgdaInductiveConstructor{◃}}\AgdaSpace{}%
\AgdaBound{P}\AgdaSpace{}%
\AgdaOperator{\AgdaFunction{$\rrbracket_{\lhd}$}}\AgdaSpace{}%
\AgdaBound{S₁}\AgdaSpace{}%
\AgdaOperator{\AgdaInductiveConstructor{◃}}\AgdaSpace{}%
\AgdaSymbol{λ}\AgdaSpace{}%
\AgdaSymbol{\{}\AgdaSpace{}%
\AgdaSymbol{(}\AgdaBound{s}\AgdaSpace{}%
\AgdaOperator{\AgdaInductiveConstructor{,}}\AgdaSpace{}%
\AgdaBound{γ}\AgdaSymbol{)}\AgdaSpace{}%
\AgdaSymbol{→}\AgdaSpace{}%
\AgdaSymbol{(}\AgdaBound{s₁}\AgdaSpace{}%
\AgdaSymbol{:}\AgdaSpace{}%
\AgdaBound{P}\AgdaSpace{}%
\AgdaBound{s}\AgdaSymbol{)}\AgdaSpace{}%
\AgdaSymbol{→}\AgdaSpace{}%
\AgdaBound{P₁}\AgdaSpace{}%
\AgdaSymbol{(}\AgdaBound{γ}\AgdaSpace{}%
\AgdaBound{s₁}\AgdaSymbol{)}\AgdaSpace{}%
\AgdaSymbol{\}}\<%
\\
\\[\AgdaEmptyExtraSkip]%
\>[0]\AgdaFunction{t}\AgdaSpace{}%
\AgdaSymbol{:}\AgdaSpace{}%
\AgdaPrimitiveType{Set}\AgdaSpace{}%
\AgdaSymbol{→}\AgdaSpace{}%
\AgdaPrimitiveType{Set}\<%
\\
\>[0]\AgdaFunction{t}\AgdaSpace{}%
\AgdaBound{X}\AgdaSpace{}%
\AgdaSymbol{=}\AgdaSpace{}%
\AgdaOperator{\AgdaFunction{⟦}}\AgdaSpace{}%
\AgdaSymbol{(}\AgdaDatatype{ℕ}\AgdaSpace{}%
\AgdaOperator{\AgdaInductiveConstructor{◃}}\AgdaSpace{}%
\AgdaDatatype{Fin}\AgdaSymbol{)}\AgdaSpace{}%
\AgdaOperator{\AgdaFunction{⋄}}\AgdaSpace{}%
\AgdaSymbol{((}\AgdaDatatype{ℕ}\AgdaSpace{}%
\AgdaOperator{\AgdaInductiveConstructor{◃}}\AgdaSpace{}%
\AgdaDatatype{Fin}\AgdaSymbol{)}\AgdaSpace{}%
\AgdaOperator{\AgdaFunction{⋄}}\AgdaSpace{}%
\AgdaSymbol{(}\AgdaDatatype{ℕ}\AgdaSpace{}%
\AgdaOperator{\AgdaInductiveConstructor{◃}}\AgdaSpace{}%
\AgdaDatatype{Fin}\AgdaSymbol{))}\AgdaSpace{}%
\AgdaOperator{\AgdaFunction{$\rrbracket_{\lhd}$}}\AgdaSpace{}%
\AgdaBound{X}\<%
\\
\\[\AgdaEmptyExtraSkip]%
\>[0]\AgdaFunction{sanity}\AgdaSpace{}%
\AgdaSymbol{:}\AgdaSpace{}%
\AgdaFunction{t}\AgdaSpace{}%
\AgdaOperator{\AgdaDatatype{≡}}\AgdaSpace{}%
\AgdaSymbol{(λ}\AgdaSpace{}%
\AgdaBound{X}\AgdaSpace{}%
\AgdaSymbol{→}\<%
\\
\>[0][@{}l@{\AgdaIndent{0}}]%
\>[2]\AgdaRecord{Σ}%
\>[176I]\AgdaSymbol{(}\<%
\\
\>[176I][@{}l@{\AgdaIndent{0}}]%
\>[7]\AgdaComment{\ {-}{-}Σ ℕ (λ s → Fin s → Σ ℕ (λ s₁ → Fin s₁ → ℕ))}\<%
\\
\>[7]\AgdaOperator{\AgdaFunction{⟦}}\AgdaSpace{}%
\AgdaDatatype{ℕ}\AgdaSpace{}%
\AgdaOperator{\AgdaInductiveConstructor{◃}}\AgdaSpace{}%
\AgdaDatatype{Fin}\AgdaSpace{}%
\AgdaOperator{\AgdaFunction{$\rrbracket_{\lhd}$}}\AgdaSpace{}%
\AgdaSymbol{(}\AgdaOperator{\AgdaFunction{⟦}}\AgdaSpace{}%
\AgdaDatatype{ℕ}\AgdaSpace{}%
\AgdaOperator{\AgdaInductiveConstructor{◃}}\AgdaSpace{}%
\AgdaDatatype{Fin}\AgdaSpace{}%
\AgdaOperator{\AgdaFunction{$\rrbracket_{\lhd}$}}\AgdaSpace{}%
\AgdaDatatype{ℕ}\AgdaSymbol{)}\<%
\\
\>[.][@{}l@{}]\<[176I]%
\>[4]\AgdaSymbol{)}\<%
\\
\>[4]\AgdaSymbol{(λ}\AgdaSpace{}%
\AgdaBound{s}\AgdaSpace{}%
\AgdaSymbol{→}%
\>[189I]\AgdaKeyword{let}\AgdaSpace{}%
\AgdaBound{ss}\AgdaSpace{}%
\AgdaOperator{\AgdaInductiveConstructor{,}}\AgdaSpace{}%
\AgdaBound{f}\AgdaSpace{}%
\AgdaSymbol{=}\AgdaSpace{}%
\AgdaBound{s}\AgdaSpace{}%
\AgdaKeyword{in}\<%
\\
\>[.][@{}l@{}]\<[189I]%
\>[11]\AgdaSymbol{((}\AgdaBound{s₁}\AgdaSpace{}%
\AgdaSymbol{:}\AgdaSpace{}%
\AgdaDatatype{Fin}\AgdaSpace{}%
\AgdaBound{ss}\AgdaSymbol{)}\AgdaSpace{}%
\AgdaSymbol{→}%
\>[200I]\AgdaKeyword{let}\AgdaSpace{}%
\AgdaBound{ss1}\AgdaSpace{}%
\AgdaOperator{\AgdaInductiveConstructor{,}}\AgdaSpace{}%
\AgdaBound{ff}\AgdaSpace{}%
\AgdaSymbol{=}\AgdaSpace{}%
\AgdaBound{f}\AgdaSpace{}%
\AgdaBound{s₁}\AgdaSpace{}%
\AgdaKeyword{in}\<%
\\
\>[.][@{}l@{}]\<[200I]%
\>[28]\AgdaSymbol{(}\AgdaBound{s₂}\AgdaSpace{}%
\AgdaSymbol{:}\AgdaSpace{}%
\AgdaDatatype{Fin}\AgdaSpace{}%
\AgdaBound{ss1}\AgdaSymbol{)}\AgdaSpace{}%
\AgdaSymbol{→}\<%
\\
\>[28]\AgdaDatatype{Fin}\AgdaSpace{}%
\AgdaSymbol{(}\AgdaBound{ff}\AgdaSpace{}%
\AgdaBound{s₂}\AgdaSymbol{))}\<%
\\
\>[11]\AgdaSymbol{→}\AgdaSpace{}%
\AgdaBound{X}\AgdaSymbol{))}\<%
\\
\>[0]\AgdaFunction{sanity}\AgdaSpace{}%
\AgdaSymbol{=}\AgdaSpace{}%
\AgdaInductiveConstructor{refl}\<%
\\
\\[\AgdaEmptyExtraSkip]%
\\[\AgdaEmptyExtraSkip]%
\>[0]\AgdaOperator{\AgdaFunction{\AgdaUnderscore{}⊕\AgdaUnderscore{}}}%
\>[217I]\AgdaSymbol{:}\AgdaSpace{}%
\AgdaSymbol{∀}\AgdaSpace{}%
\AgdaSymbol{\{}\AgdaBound{a}\AgdaSpace{}%
\AgdaBound{b}\AgdaSymbol{\}\{}\AgdaBound{A}\AgdaSpace{}%
\AgdaBound{B}\AgdaSpace{}%
\AgdaSymbol{:}\AgdaSpace{}%
\AgdaPrimitiveType{Set}\AgdaSpace{}%
\AgdaBound{a}\AgdaSymbol{\}\{}\AgdaBound{C}\AgdaSpace{}%
\AgdaSymbol{:}\AgdaSpace{}%
\AgdaBound{A}\AgdaSpace{}%
\AgdaOperator{\AgdaDatatype{⊎}}\AgdaSpace{}%
\AgdaBound{B}\AgdaSpace{}%
\AgdaSymbol{→}\AgdaSpace{}%
\AgdaPrimitiveType{Set}\AgdaSpace{}%
\AgdaBound{b}\AgdaSymbol{\}}\<%
\\
\>[.][@{}l@{}]\<[217I]%
\>[4]\AgdaSymbol{→}\AgdaSpace{}%
\AgdaSymbol{((}\AgdaBound{x}\AgdaSpace{}%
\AgdaSymbol{:}\AgdaSpace{}%
\AgdaBound{A}\AgdaSymbol{)}\AgdaSpace{}%
\AgdaSymbol{→}\AgdaSpace{}%
\AgdaBound{C}\AgdaSpace{}%
\AgdaSymbol{(}\AgdaInductiveConstructor{inj₁}\AgdaSpace{}%
\AgdaBound{x}\AgdaSymbol{))}\<%
\\
\>[4]\AgdaSymbol{→}\AgdaSpace{}%
\AgdaSymbol{((}\AgdaBound{x}\AgdaSpace{}%
\AgdaSymbol{:}\AgdaSpace{}%
\AgdaBound{B}\AgdaSymbol{)}\AgdaSpace{}%
\AgdaSymbol{→}\AgdaSpace{}%
\AgdaBound{C}\AgdaSpace{}%
\AgdaSymbol{(}\AgdaInductiveConstructor{inj₂}\AgdaSpace{}%
\AgdaBound{x}\AgdaSymbol{))}\<%
\\
\>[4]\AgdaSymbol{→}\AgdaSpace{}%
\AgdaSymbol{(}\AgdaBound{x}\AgdaSpace{}%
\AgdaSymbol{:}\AgdaSpace{}%
\AgdaBound{A}\AgdaSpace{}%
\AgdaOperator{\AgdaDatatype{⊎}}\AgdaSpace{}%
\AgdaBound{B}\AgdaSymbol{)}\<%
\\
\>[4]\AgdaSymbol{→}\AgdaSpace{}%
\AgdaBound{C}\AgdaSpace{}%
\AgdaBound{x}\<%
\\
\>[0]\AgdaSymbol{(}\AgdaBound{f}\AgdaSpace{}%
\AgdaOperator{\AgdaFunction{⊕}}\AgdaSpace{}%
\AgdaBound{g}\AgdaSymbol{)}\AgdaSpace{}%
\AgdaSymbol{(}\AgdaInductiveConstructor{inj₁}\AgdaSpace{}%
\AgdaBound{x}\AgdaSymbol{)}\AgdaSpace{}%
\AgdaSymbol{=}\AgdaSpace{}%
\AgdaBound{f}\AgdaSpace{}%
\AgdaBound{x}\<%
\\
\>[0]\AgdaSymbol{(}\AgdaBound{f}\AgdaSpace{}%
\AgdaOperator{\AgdaFunction{⊕}}\AgdaSpace{}%
\AgdaBound{g}\AgdaSymbol{)}\AgdaSpace{}%
\AgdaSymbol{(}\AgdaInductiveConstructor{inj₂}\AgdaSpace{}%
\AgdaBound{y}\AgdaSymbol{)}\AgdaSpace{}%
\AgdaSymbol{=}\AgdaSpace{}%
\AgdaBound{g}\AgdaSpace{}%
\AgdaBound{y}\<%
\\
\\[\AgdaEmptyExtraSkip]%
\>[0]\AgdaFunction{plus}\AgdaSpace{}%
\AgdaSymbol{:}\AgdaSpace{}%
\AgdaRecord{Con}\AgdaSpace{}%
\AgdaSymbol{→}\AgdaSpace{}%
\AgdaRecord{Con}\AgdaSpace{}%
\AgdaSymbol{→}\AgdaSpace{}%
\AgdaRecord{Con}\<%
\\
\>[0]\AgdaFunction{plus}\AgdaSpace{}%
\AgdaSymbol{(}\AgdaBound{A}\AgdaSpace{}%
\AgdaOperator{\AgdaInductiveConstructor{◃}}\AgdaSpace{}%
\AgdaBound{B}\AgdaSymbol{)}\AgdaSpace{}%
\AgdaSymbol{(}\AgdaBound{C}\AgdaSpace{}%
\AgdaOperator{\AgdaInductiveConstructor{◃}}\AgdaSpace{}%
\AgdaBound{D}\AgdaSymbol{)}\AgdaSpace{}%
\AgdaSymbol{=}\AgdaSpace{}%
\AgdaSymbol{(}\AgdaBound{A}\AgdaSpace{}%
\AgdaOperator{\AgdaDatatype{⊎}}\AgdaSpace{}%
\AgdaBound{C}\AgdaSymbol{)}\AgdaSpace{}%
\AgdaOperator{\AgdaInductiveConstructor{◃}}\AgdaSpace{}%
\AgdaSymbol{(}\AgdaBound{B}\AgdaSpace{}%
\AgdaOperator{\AgdaFunction{⊕}}\AgdaSpace{}%
\AgdaBound{D}\AgdaSymbol{)}\<%
\\
\\[\AgdaEmptyExtraSkip]%
\>[0]\AgdaFunction{inner}\AgdaSpace{}%
\AgdaSymbol{:}\AgdaSpace{}%
\AgdaDatatype{Fin}\AgdaSpace{}%
\AgdaNumber{2}\AgdaSpace{}%
\AgdaSymbol{→}\AgdaSpace{}%
\AgdaRecord{Σ}\AgdaSpace{}%
\AgdaDatatype{ℕ}\AgdaSpace{}%
\AgdaSymbol{λ}\AgdaSpace{}%
\AgdaBound{n}\AgdaSpace{}%
\AgdaSymbol{→}\AgdaSpace{}%
\AgdaDatatype{Fin}\AgdaSpace{}%
\AgdaBound{n}\AgdaSpace{}%
\AgdaSymbol{→}\AgdaSpace{}%
\AgdaDatatype{ℕ}\<%
\\
\>[0]\AgdaFunction{inner}\AgdaSpace{}%
\AgdaInductiveConstructor{zero}%
\>[17]\AgdaSymbol{=}\AgdaSpace{}%
\AgdaNumber{1}\AgdaSpace{}%
\AgdaOperator{\AgdaInductiveConstructor{,}}\AgdaSpace{}%
\AgdaSymbol{λ}\AgdaSpace{}%
\AgdaBound{\AgdaUnderscore{}}\AgdaSpace{}%
\AgdaSymbol{→}\AgdaSpace{}%
\AgdaNumber{3}\<%
\\
\>[0]\AgdaFunction{inner}\AgdaSpace{}%
\AgdaSymbol{(}\AgdaInductiveConstructor{suc}\AgdaSpace{}%
\AgdaInductiveConstructor{zero}\AgdaSymbol{)}\AgdaSpace{}%
\AgdaSymbol{=}\AgdaSpace{}%
\AgdaNumber{2}\AgdaSpace{}%
\AgdaOperator{\AgdaInductiveConstructor{,}}\AgdaSpace{}%
\AgdaSymbol{λ}\AgdaSpace{}%
\AgdaBound{\AgdaUnderscore{}}\AgdaSpace{}%
\AgdaSymbol{→}\AgdaSpace{}%
\AgdaNumber{4}\<%
\\
\\[\AgdaEmptyExtraSkip]%
\>[0]\AgdaFunction{idx-t}\AgdaSpace{}%
\AgdaSymbol{=}%
\>[317I]\AgdaSymbol{((}\AgdaBound{i}\AgdaSpace{}%
\AgdaSymbol{:}\AgdaSpace{}%
\AgdaDatatype{Fin}\AgdaSpace{}%
\AgdaNumber{2}\AgdaSymbol{)}\<%
\\
\>[317I][@{}l@{\AgdaIndent{0}}]%
\>[9]\AgdaSymbol{(}\AgdaBound{j}\AgdaSpace{}%
\AgdaSymbol{:}\AgdaSpace{}%
\AgdaDatatype{Fin}\AgdaSpace{}%
\AgdaSymbol{(}\AgdaField{proj₁}\AgdaSpace{}%
\AgdaSymbol{(}\AgdaFunction{inner}\AgdaSpace{}%
\AgdaBound{i}\AgdaSymbol{)))}\AgdaSpace{}%
\AgdaSymbol{→}\<%
\\
\>[9]\AgdaDatatype{Fin}\AgdaSpace{}%
\AgdaSymbol{(}\AgdaField{proj₂}\AgdaSpace{}%
\AgdaSymbol{(}\AgdaFunction{inner}\AgdaSpace{}%
\AgdaBound{i}\AgdaSymbol{)}\AgdaSpace{}%
\AgdaBound{j}\AgdaSymbol{))}\AgdaSpace{}%
\AgdaSymbol{→}\<%
\\
\>[.][@{}l@{}]\<[317I]%
\>[8]\AgdaDatatype{ℕ}\<%
\\
\\[\AgdaEmptyExtraSkip]%
\>[0]\AgdaComment{\ {-}{-} The actual array.}\<%
\\
\>[0]\AgdaFunction{ar-rhs}\AgdaSpace{}%
\AgdaSymbol{:}\AgdaSpace{}%
\AgdaFunction{t}\AgdaSpace{}%
\AgdaDatatype{ℕ}\<%
\\
\>[0]\AgdaFunction{ar-rhs}\AgdaSpace{}%
\AgdaSymbol{=}\AgdaSpace{}%
\AgdaSymbol{(}\AgdaNumber{2}\AgdaSpace{}%
\AgdaOperator{\AgdaInductiveConstructor{,}}\AgdaSpace{}%
\AgdaFunction{inner}\AgdaSymbol{)}\AgdaSpace{}%
\AgdaOperator{\AgdaInductiveConstructor{,}}\AgdaSpace{}%
\AgdaSymbol{λ}\AgdaSpace{}%
\AgdaBound{\AgdaUnderscore{}}\AgdaSpace{}%
\AgdaSymbol{→}\AgdaSpace{}%
\AgdaNumber{42}\<%
\\
\>[0]\<%
\end{code}
The shape of the tree is given by a tree that does not store any
data in its nodes.  Positions into this tree are all the valid paths
leading from the root to some node in the shape tree.

For further details on numerous container properties refer
to~\cite{containers,containers1,containers2,containers3}.

\subsection{Rectangular Arrays}
Now let us explore how containers can be used to define rectangular
multi-dimensional arrays.
The shape of a $d$-dimensional rectangular array can be represented as a
$d$-element tuple of natural numbers.  For a tuple $(s_1, \dots, s_d)$,
every value $s_i$ represents the number of elements over the axis $i$ in
the array with that shape.  All the array elements are of type $X$:
\[
    Array\ X = \prodt{d: \Nat}
                     {\prodt{s: Fin\ d \to \Nat}
                            {\left(\left(\prodt{i: Fin\ d}{Fin\ (s\ i)}\right)
                             \to X\right)}}
\]
where $d$ is the number of dimensions, $s$ is a $d$-element tuple of \Nat{}
that we model as a function, and the content of the array is a function from
indices to values of type $X$.  Indices are $d$-element tuples of natural
numbers where every $i$-th element is bound by the corresponding position
in the shape vector $s\ i$.


\paragraph{Properties}
The type of indices ensures that
out-of-bound access is not possible.  Arrays with empty
shapes inhabit the $Array$ type as there exists a function from
$Fin\ 0$ (empty set) to \Nat{}.
Arrays of this kind are often called scalars.
\[
    s: Fin\ 0 \to \Nat
    \qquad \prodt{i:Fin\ 0}{Fin\ (s\ i)} \cong \top
    \qquad X \cong Array\ X\ 0\ s\ (f: \top \to X)
\]
Most array processing languages treat scalars as degenerate
(0-dimensional) arrays.

By similar reasoning, the $Array$ type allows an infinite number of empty
arrays (arrays with no elements).  An empty array can be characterised by
a shape function that evaluates to zero at one of its inputs.
(There are no indices permitted in that dimension.) 
\[
    \exists (i: Fin\ d)\ \ s\ i = 0
    \implies \prodt{i: Fin\ d}{Fin\ (s\ i)} \cong \bot
\]
Empty arrays do not contain any elements, but they do exist, due to existence
of a function of type $\bot \to X$, whatever the type $X$.  Empty arrays are often found useful in
practice, for example as neutral elements for array concatenations.

\subsection{The $\diamond$ Operation}
Let us find a container formulation for the $Array$ type.  To do so, we
uncurry the first two arguments $d$ and $s$ as follows:
\[
    Array\ X = \contint{ \left(\sumt{d: \Nat}{Fin\ d \to \Nat}\right)
                      \lhd \lambda (d, s) \to \prodt{i: Fin\ d}{Fin\ (s\ i)}
                    }\ X
\]
We can notice that the first sigma can be represented as a container as well:
\[
    \sumt{d: \Nat}{Fin\ d \to \Nat}
    = \contint{\Nat \lhd Fin}\Nat
\]
Using this observation let us generalise $Array$ as a result of the following
container operation:
\[
    (A \lhd B) \diamond (C \lhd D)
    = \contint{A \lhd B}C \lhd \lambda (a, s) \to \prodt{B\ a}{D \circ s}
\]
Using $\diamond$ we describe homogeneous rectangular arrays as:
\[
    Array\ X = \contint{(\Nat \lhd Fin) \diamond (\Nat \lhd Fin)}\ X
\]

\paragraph{Discussion}
One may think about the $\diamond$ operation as of a generalisation of the
container operation that is often referred as Hancock's tensor.
The tensor operation on containers is defined as:
\[
     (A \lhd B) \otimes (C \lhd D)
        = (A \times C) \lhd \lambda (a,c) \to B\ a \times D\ c
\]
If we want to compute an $n$-fold tensor product of the container
$C \lhd D$:
\[
     (C \lhd D) \otimes (C \lhd D)\otimes (C \lhd D)\otimes \cdots
\]
we need to specify the boundaries of the product.  Instead of giving
a number, we can use another container, and use its index-space to
encode the count.  One might write:
\[
        (A \lhd B) \diamond (C \lhd D) =
        \bigotimes \contint{A \lhd B} (C \lhd D)
\]
We ``set the bounds'' of the iterated tensor product using a ``count
container'' $A \lhd B$.  This gives us a family of containers
$(a: A, f: B\ a \to (C \lhd D))$ and we compute tensor product
of all the elements produced by $f$.


Further, we notice the following analogy. 
Similarly to the way $\otimes$
replaces\footnote{We have overloaded + and $\times$ for $B$ and $D$.}
$+$ with $\times$ in the container coproduct:
\begin{align*}
        (A \lhd B) + (C \lhd D) &= (A + C) \lhd (B + D)\\
        (A \lhd B) \otimes (C \lhd D) &= (A \times C) \lhd (B \times D)
\end{align*}
in the same way $\diamond$ replaces $\sum$ with $\prod$ in the container
composition:
\begin{align*}
        (A \lhd B) \circ (C \lhd D)
           &= \contint{A \lhd B}C \lhd
              \lambda (a, \gamma) \to \sumt{B\ a} D \circ \gamma\\
        (A \lhd B) \diamond (C \lhd D)
           &= \contint{A \lhd B}C \lhd
              \lambda (a, \gamma) \to \prodt{B\ a} D \circ \gamma
\end{align*}
%
%
%
\paragraph{Iteration}
Let us now explore iterated applications of $-\diamond\,(\Nat \lhd Fin)$ and
$(\Nat \lhd Fin)\,\diamond-$ treating $\diamond$ first as a as left-associative and
then as a right-associative operation.

\begin{equation*}
\begin{split}
    \left(
        \left(\Nat \lhd Fin\right) \diamond \left(\Nat \lhd Fin\right)
    \right)
    \diamond \left(\Nat \lhd Fin\right)
  &= \left(
      \contint{\Nat \lhd Fin}\Nat
      \lhd \lambda (d, s) \to \prod_{Fin\ d} (Fin \circ s)
    \right)
    \diamond \left(\Nat \lhd Fin\right) \\
  &= \contint{
      \contint{\Nat \lhd Fin}\Nat
      \lhd \lambda (d, s) \to \prod_{Fin\ d} (Fin \circ s)
    }\Nat\\
  &\qquad
    \lhd \lambda ((d, s), v)
         \to \prod\nolimits_{\textstyle\prod_{Fin\ d}(Fin \circ s)} (Fin \circ v)
\end{split}
\end{equation*}

In this left-associative case we see that we obtain a generalisation
of multi-dimensional arrays which we call level-3 arrays.  The shape of
a level-2 array is given by a pair $(d, s)$, where $d$ is
the dimensionality and $s$ is the shape.  The indices into level-2
arrays are vectors of $Fin$-s of the same length as $s$.  The shape of a
level-3 array is a level-2 array $v$ of \Nat.  The
indices into such an array are level-2 arrays of $Fin$-s with the same
shape as $v$.  We can get even higher levels by further
application of $-\diamond\,(\Nat \lhd Fin)$.

When $\diamond$ is right-associative, \ie{} we
successively apply $(\Nat \lhd Fin)\,\diamond-$ (on the left) we get:
\begin{equation*}
\begin{split}
    \left(\Nat \lhd Fin\right)
    \diamond
    \left(
        \left(\Nat \lhd Fin\right) \diamond \left(\Nat \lhd Fin\right)
    \right)
  &=\left(\Nat \lhd Fin\right)
    \diamond
    \left(
      \contint{\Nat \lhd Fin}\Nat
      \lhd \lambda (d, s) \to \prod_{Fin\ d} (Fin \circ s)
    \right)\\
  &=\contint{\Nat \lhd Fin}\left(\contint{\Nat \lhd Fin}\Nat\right)\\
  &\quad\lhd
    \lambda (m, f)
         \to \prod_{Fin\ m} \left(\lambda (n, s) \to \prod_{Fin\ n} Fin \circ s\right)
         \circ f
\end{split}
\end{equation*}

Note that this is well-formed because:
\[
    f: Fin\ m \to \sum_{n:\Nat} (Fin\ n \to \Nat)
\]
The shape of such an array is a vector of vectors of \Nat.  The shape is a
2-dimensional array, but it does not have a rectangular structure, as its rows
could be of different lengths.  The array itself still has a rectangular
structure and is indexed by the $Fin\ x$ ``tuples'' where $x$
iterates over the elements of the shape.  For example, we may create a shape
\(
\begin{pmatrix}
        3 &\\
        4 & 5
\end{pmatrix}
\)
that is encoded with the type
\(%
    \sumt{m}{\left({Fin\ m \to \sumt{n}
                    {\left({Fin\ n \to \Nat}\right)}}
             \right)}
\)
where $m = 2$ and the corresponding two sigmas are $(1, \lambda\ \_ \to 3)$ and
$(2, \lambda\ \{0 \to 4; 1 \to 5\})$.  The array itself would be isomorphic to
the 3-d array of shape $3\times4\times5$.  Further applications of $(\Nat \lhd Fin)\,
\diamond-$ (on the left) will turn shapes into 3-d irregular arrays, as it simply
generates a vector of the shapes from the previous level.

This difference is somewhat natural if we recall the explanation of the $\diamond$
operation via ``count containers'' and tensor product.  Left application of $\diamond$
acts on the original ``count container'' enriching its structure.  The right
application turns the structure obtained at the level $l$ into the count
container for the level $l+1$.


\section{Array Levels}
\begin{code}[hide]%
\>[0]\AgdaKeyword{open}\AgdaSpace{}%
\AgdaKeyword{import}\AgdaSpace{}%
\AgdaModule{Data.Product}\<%
\\
\>[0]\AgdaKeyword{open}\AgdaSpace{}%
\AgdaKeyword{import}\AgdaSpace{}%
\AgdaModule{Data.Nat}\<%
\\
\>[0]\AgdaKeyword{open}\AgdaSpace{}%
\AgdaKeyword{import}\AgdaSpace{}%
\AgdaModule{Data.Unit}\<%
\\
\>[0]\AgdaKeyword{open}\AgdaSpace{}%
\AgdaKeyword{import}\AgdaSpace{}%
\AgdaModule{Data.Empty}\<%
\\
\>[0]\AgdaKeyword{open}\AgdaSpace{}%
\AgdaKeyword{import}\AgdaSpace{}%
\AgdaModule{Data.Fin}\AgdaSpace{}%
\AgdaKeyword{hiding}\AgdaSpace{}%
\AgdaSymbol{(}\AgdaOperator{\AgdaFunction{\AgdaUnderscore{}+\AgdaUnderscore{}}}\AgdaSymbol{)}\<%
\\
\>[0]\AgdaKeyword{open}\AgdaSpace{}%
\AgdaKeyword{import}\AgdaSpace{}%
\AgdaModule{Relation.Binary.PropositionalEquality}\<%
\\
\\[\AgdaEmptyExtraSkip]%
\>[0]\AgdaKeyword{open}\AgdaSpace{}%
\AgdaKeyword{import}\AgdaSpace{}%
\AgdaModule{Function}\<%
\\
\\[\AgdaEmptyExtraSkip]%
\>[0]\AgdaKeyword{record}\AgdaSpace{}%
\AgdaRecord{Con}\AgdaSpace{}%
\AgdaSymbol{:}\AgdaSpace{}%
\AgdaPrimitiveType{Set₁}\AgdaSpace{}%
\AgdaKeyword{where}\<%
\\
\>[0][@{}l@{\AgdaIndent{0}}]%
\>[2]\AgdaKeyword{constructor}\AgdaSpace{}%
\AgdaOperator{\AgdaInductiveConstructor{\AgdaUnderscore{}◃\AgdaUnderscore{}}}\<%
\\
\>[2]\AgdaKeyword{field}\<%
\\
\>[2][@{}l@{\AgdaIndent{0}}]%
\>[4]\AgdaField{Sh}\AgdaSpace{}%
\AgdaSymbol{:}\AgdaSpace{}%
\AgdaPrimitiveType{Set}\<%
\\
\>[4]\AgdaField{Po}\AgdaSpace{}%
\AgdaSymbol{:}\AgdaSpace{}%
\AgdaBound{Sh}\AgdaSpace{}%
\AgdaSymbol{→}\AgdaSpace{}%
\AgdaPrimitiveType{Set}\<%
\\
\>[2]\AgdaOperator{\AgdaFunction{⟦\AgdaUnderscore{}$\rrbracket_{\lhd}$}}%
\>[8]\AgdaSymbol{:}\AgdaSpace{}%
\AgdaPrimitiveType{Set}\AgdaSpace{}%
\AgdaSymbol{→}\AgdaSpace{}%
\AgdaPrimitiveType{Set}\<%
\\
\>[2]\AgdaOperator{\AgdaFunction{⟦\AgdaUnderscore{}$\rrbracket_{\lhd}$}}\AgdaSpace{}%
\AgdaBound{X}\AgdaSpace{}%
\AgdaSymbol{=}\AgdaSpace{}%
\AgdaRecord{Σ}\AgdaSpace{}%
\AgdaField{Sh}\AgdaSpace{}%
\AgdaSymbol{λ}\AgdaSpace{}%
\AgdaBound{s}\AgdaSpace{}%
\AgdaSymbol{→}\AgdaSpace{}%
\AgdaField{Po}\AgdaSpace{}%
\AgdaBound{s}\AgdaSpace{}%
\AgdaSymbol{→}\AgdaSpace{}%
\AgdaBound{X}\<%
\\
\>[0]\AgdaKeyword{open}\AgdaSpace{}%
\AgdaModule{Con}\AgdaSpace{}%
\AgdaKeyword{public}\<%
\\
\>[0]\AgdaKeyword{infixr}\AgdaSpace{}%
\AgdaNumber{1}\AgdaSpace{}%
\AgdaOperator{\AgdaInductiveConstructor{\AgdaUnderscore{}◃\AgdaUnderscore{}}}\<%
\end{code}

With the $\diamond$ operation in hand, we can define a hierarchy of arrays
with levels in the following way.
\begin{code}%
\>[0]\AgdaOperator{\AgdaFunction{\AgdaUnderscore{}⋄\AgdaUnderscore{}}}\AgdaSpace{}%
\AgdaSymbol{:}\AgdaSpace{}%
\AgdaRecord{Con}\AgdaSpace{}%
\AgdaSymbol{→}\AgdaSpace{}%
\AgdaRecord{Con}\AgdaSpace{}%
\AgdaSymbol{→}\AgdaSpace{}%
\AgdaRecord{Con}\<%
\\
\>[0]\AgdaSymbol{(}\AgdaBound{S}\AgdaSpace{}%
\AgdaOperator{\AgdaInductiveConstructor{◃}}\AgdaSpace{}%
\AgdaBound{P}\AgdaSymbol{)}\AgdaSpace{}%
\AgdaOperator{\AgdaFunction{⋄}}\AgdaSpace{}%
\AgdaSymbol{(}\AgdaBound{S₁}\AgdaSpace{}%
\AgdaOperator{\AgdaInductiveConstructor{◃}}\AgdaSpace{}%
\AgdaBound{P₁}\AgdaSymbol{)}\AgdaSpace{}%
\AgdaSymbol{=}\AgdaSpace{}%
\AgdaOperator{\AgdaFunction{⟦}}\AgdaSpace{}%
\AgdaBound{S}\AgdaSpace{}%
\AgdaOperator{\AgdaInductiveConstructor{◃}}\AgdaSpace{}%
\AgdaBound{P}\AgdaSpace{}%
\AgdaOperator{\AgdaFunction{$\rrbracket_{\lhd}$}}\AgdaSpace{}%
\AgdaBound{S₁}\AgdaSpace{}%
\AgdaOperator{\AgdaInductiveConstructor{◃}}\AgdaSpace{}%
\AgdaSymbol{λ}\AgdaSpace{}%
\AgdaSymbol{\{}\AgdaSpace{}%
\AgdaSymbol{(}\AgdaBound{s}\AgdaSpace{}%
\AgdaOperator{\AgdaInductiveConstructor{,}}\AgdaSpace{}%
\AgdaBound{γ}\AgdaSymbol{)}\AgdaSpace{}%
\AgdaSymbol{→}\AgdaSpace{}%
\AgdaSymbol{(}\AgdaBound{s₁}\AgdaSpace{}%
\AgdaSymbol{:}\AgdaSpace{}%
\AgdaBound{P}\AgdaSpace{}%
\AgdaBound{s}\AgdaSymbol{)}\AgdaSpace{}%
\AgdaSymbol{→}\AgdaSpace{}%
\AgdaBound{P₁}\AgdaSpace{}%
\AgdaSymbol{(}\AgdaBound{γ}\AgdaSpace{}%
\AgdaBound{s₁}\AgdaSymbol{)}\AgdaSpace{}%
\AgdaSymbol{\}}\<%
\\
\\[\AgdaEmptyExtraSkip]%
\>[0]\AgdaFunction{A}\AgdaSpace{}%
\AgdaSymbol{:}\AgdaSpace{}%
\AgdaDatatype{ℕ}\AgdaSpace{}%
\AgdaSymbol{→}\AgdaSpace{}%
\AgdaRecord{Con}\<%
\\
\>[0]\AgdaFunction{A}\AgdaSpace{}%
\AgdaInductiveConstructor{zero}%
\>[10]\AgdaSymbol{=}\AgdaSpace{}%
\AgdaRecord{⊤}\AgdaSpace{}%
\AgdaOperator{\AgdaInductiveConstructor{◃}}\AgdaSpace{}%
\AgdaSymbol{λ}\AgdaSpace{}%
\AgdaBound{\AgdaUnderscore{}}\AgdaSpace{}%
\AgdaSymbol{→}\AgdaSpace{}%
\AgdaRecord{⊤}\<%
\\
\>[0]\AgdaFunction{A}\AgdaSpace{}%
\AgdaSymbol{(}\AgdaInductiveConstructor{suc}\AgdaSpace{}%
\AgdaBound{x}\AgdaSymbol{)}\AgdaSpace{}%
\AgdaSymbol{=}\AgdaSpace{}%
\AgdaSymbol{(}\AgdaFunction{A}\AgdaSpace{}%
\AgdaBound{x}\AgdaSymbol{)}\AgdaSpace{}%
\AgdaOperator{\AgdaFunction{⋄}}\AgdaSpace{}%
\AgdaSymbol{(}\AgdaDatatype{ℕ}\AgdaSpace{}%
\AgdaOperator{\AgdaInductiveConstructor{◃}}\AgdaSpace{}%
\AgdaDatatype{Fin}\AgdaSymbol{)}\<%
\end{code}
$\contint{A\ n}$ is a level-$n$ array.  
Our iteration begins with 
level-0 arrays, where all the shapes are singletons.
Level-0 arrays are often referred
to as scalars in array calculi.  Even though $\contint{A\ 0}\ X$ is isomorphic
to $X$, it still makes sense to have both: 
a level-polymorphic array
operation is applicable to scalars and is not applicable to $X$.

Unfortunately, this data structure is not very convenient for observing shape
relations in 
array operations.  Consider a regular \texttt{cons}
operator on a level-1 array.
\begin{code}%
\>[0]\AgdaFunction{cons}\AgdaSpace{}%
\AgdaSymbol{:}\AgdaSpace{}%
\AgdaSymbol{∀}\AgdaSpace{}%
\AgdaSymbol{\{}\AgdaBound{X}\AgdaSymbol{\}}\AgdaSpace{}%
\AgdaSymbol{→}\AgdaSpace{}%
\AgdaBound{X}\AgdaSpace{}%
\AgdaSymbol{→}\AgdaSpace{}%
\AgdaOperator{\AgdaFunction{⟦}}\AgdaSpace{}%
\AgdaFunction{A}\AgdaSpace{}%
\AgdaNumber{1}\AgdaSpace{}%
\AgdaOperator{\AgdaFunction{$\rrbracket_{\lhd}$}}\AgdaSpace{}%
\AgdaBound{X}\AgdaSpace{}%
\AgdaSymbol{→}\AgdaSpace{}%
\AgdaOperator{\AgdaFunction{⟦}}\AgdaSpace{}%
\AgdaFunction{A}\AgdaSpace{}%
\AgdaNumber{1}\AgdaSpace{}%
\AgdaOperator{\AgdaFunction{$\rrbracket_{\lhd}$}}\AgdaSpace{}%
\AgdaBound{X}\<%
\\
\>[0]\AgdaFunction{cons}\AgdaSpace{}%
\AgdaSymbol{\{}\AgdaBound{X}\AgdaSymbol{\}}\AgdaSpace{}%
\AgdaBound{x}\AgdaSpace{}%
\AgdaSymbol{((\AgdaUnderscore{}}\AgdaSpace{}%
\AgdaOperator{\AgdaInductiveConstructor{,}}\AgdaSpace{}%
\AgdaBound{s}\AgdaSymbol{)}\AgdaSpace{}%
\AgdaOperator{\AgdaInductiveConstructor{,}}\AgdaSpace{}%
\AgdaBound{p}\AgdaSymbol{)}\AgdaSpace{}%
\AgdaSymbol{=}\AgdaSpace{}%
\AgdaSymbol{(}\AgdaInductiveConstructor{tt}\AgdaSpace{}%
\AgdaOperator{\AgdaInductiveConstructor{,}}\AgdaSpace{}%
\AgdaInductiveConstructor{suc}\AgdaSpace{}%
\AgdaOperator{\AgdaFunction{∘}}\AgdaSpace{}%
\AgdaBound{s}\AgdaSymbol{)}\AgdaSpace{}%
\AgdaOperator{\AgdaInductiveConstructor{,}}\AgdaSpace{}%
\AgdaFunction{ix-val}\AgdaSpace{}%
\AgdaKeyword{where}\<%
\\
\>[0][@{}l@{\AgdaIndent{0}}]%
\>[4]\AgdaFunction{ix-val}\AgdaSpace{}%
\AgdaSymbol{:}\AgdaSpace{}%
\AgdaSymbol{((}\AgdaBound{x}\AgdaSpace{}%
\AgdaSymbol{:}\AgdaSpace{}%
\AgdaRecord{⊤}\AgdaSymbol{)}\AgdaSpace{}%
\AgdaSymbol{→}\AgdaSpace{}%
\AgdaDatatype{Fin}\AgdaSpace{}%
\AgdaSymbol{(}\AgdaInductiveConstructor{suc}\AgdaSpace{}%
\AgdaSymbol{(}\AgdaBound{s}\AgdaSpace{}%
\AgdaBound{x}\AgdaSymbol{)))}\AgdaSpace{}%
\AgdaSymbol{→}\AgdaSpace{}%
\AgdaBound{X}\<%
\\
\>[4]\AgdaFunction{ix-val}\AgdaSpace{}%
\AgdaBound{iv}\AgdaSpace{}%
\AgdaKeyword{with}\AgdaSpace{}%
\AgdaBound{iv}\AgdaSpace{}%
\AgdaInductiveConstructor{tt}\<%
\\
\>[4]\AgdaSymbol{...}\AgdaSpace{}%
\AgdaSymbol{|}\AgdaSpace{}%
\AgdaInductiveConstructor{zero}%
\>[18]\AgdaSymbol{=}\AgdaSpace{}%
\AgdaBound{x}\<%
\\
\>[4]\AgdaSymbol{...}\AgdaSpace{}%
\AgdaSymbol{|}\AgdaSpace{}%
\AgdaSymbol{(}\AgdaInductiveConstructor{suc}\AgdaSpace{}%
\AgdaBound{j}\AgdaSymbol{)}\AgdaSpace{}%
\AgdaSymbol{=}\AgdaSpace{}%
\AgdaBound{p}\AgdaSpace{}%
\AgdaSymbol{λ}\AgdaSpace{}%
\AgdaBound{\AgdaUnderscore{}}\AgdaSpace{}%
\AgdaSymbol{→}\AgdaSpace{}%
\AgdaBound{j}\<%
\end{code}
We would like to observe from the type signature that the resulting array
is one element longer than the input.  We can surely encode this information
as follows:
\begin{code}%
\>[0]\AgdaFunction{cons+inv}\AgdaSpace{}%
\AgdaSymbol{:}\AgdaSpace{}%
\AgdaSymbol{∀}\AgdaSpace{}%
\AgdaSymbol{\{}\AgdaBound{X}\AgdaSymbol{\}}%
\>[163I]\AgdaSymbol{→}\AgdaSpace{}%
\AgdaBound{X}\AgdaSpace{}%
\AgdaSymbol{→}\AgdaSpace{}%
\AgdaSymbol{(}\AgdaBound{v}\AgdaSpace{}%
\AgdaSymbol{:}\AgdaSpace{}%
\AgdaOperator{\AgdaFunction{⟦}}\AgdaSpace{}%
\AgdaFunction{A}\AgdaSpace{}%
\AgdaNumber{1}\AgdaSpace{}%
\AgdaOperator{\AgdaFunction{$\rrbracket_{\lhd}$}}\AgdaSpace{}%
\AgdaBound{X}\AgdaSymbol{)}\<%
\\
\>[.][@{}l@{}]\<[163I]%
\>[17]\AgdaSymbol{→}%
\>[173I]\AgdaRecord{Σ}\AgdaSpace{}%
\AgdaSymbol{(}\AgdaOperator{\AgdaFunction{⟦}}\AgdaSpace{}%
\AgdaFunction{A}\AgdaSpace{}%
\AgdaNumber{1}\AgdaSpace{}%
\AgdaOperator{\AgdaFunction{$\rrbracket_{\lhd}$}}\AgdaSpace{}%
\AgdaBound{X}\AgdaSymbol{)}\<%
\\
\>[.][@{}l@{}]\<[173I]%
\>[19]\AgdaSymbol{λ}\AgdaSpace{}%
\AgdaBound{r}\AgdaSpace{}%
\AgdaSymbol{→}\AgdaSpace{}%
\AgdaField{proj₂}\AgdaSpace{}%
\AgdaSymbol{(}\AgdaField{proj₁}\AgdaSpace{}%
\AgdaBound{r}\AgdaSymbol{)}\AgdaSpace{}%
\AgdaInductiveConstructor{tt}\AgdaSpace{}%
\AgdaOperator{\AgdaDatatype{≡}}\AgdaSpace{}%
\AgdaNumber{1}\AgdaSpace{}%
\AgdaOperator{\AgdaPrimitive{+}}\AgdaSpace{}%
\AgdaField{proj₂}\AgdaSpace{}%
\AgdaSymbol{(}\AgdaField{proj₁}\AgdaSpace{}%
\AgdaBound{v}\AgdaSymbol{)}\AgdaSpace{}%
\AgdaInductiveConstructor{tt}\<%
\\
\>[0]\AgdaFunction{cons+inv}\AgdaSpace{}%
\AgdaBound{x}\AgdaSpace{}%
\AgdaBound{s}\AgdaSpace{}%
\AgdaSymbol{=}\AgdaSpace{}%
\AgdaFunction{cons}\AgdaSpace{}%
\AgdaBound{x}\AgdaSpace{}%
\AgdaBound{s}\AgdaSpace{}%
\AgdaOperator{\AgdaInductiveConstructor{,}}\AgdaSpace{}%
\AgdaInductiveConstructor{refl}\<%
\end{code}
However, as this is a frequent case, 
we find it more natural
to break the structure of the container apart, and lift the shape information
into the type.  We end up with the following array type.

\begin{code}%
\>[0]\AgdaKeyword{data}\AgdaSpace{}%
\AgdaDatatype{Ar}\AgdaSpace{}%
\AgdaSymbol{\{}\AgdaBound{a}\AgdaSymbol{\}}\AgdaSpace{}%
\AgdaSymbol{(}\AgdaBound{l}\AgdaSpace{}%
\AgdaSymbol{:}\AgdaSpace{}%
\AgdaDatatype{ℕ}\AgdaSymbol{)}\AgdaSpace{}%
\AgdaSymbol{(}\AgdaBound{X}\AgdaSpace{}%
\AgdaSymbol{:}\AgdaSpace{}%
\AgdaPrimitiveType{Set}\AgdaSpace{}%
\AgdaBound{a}\AgdaSymbol{)}\AgdaSpace{}%
\AgdaSymbol{(}\AgdaBound{s}\AgdaSpace{}%
\AgdaSymbol{:}\AgdaSpace{}%
\AgdaField{Sh}\AgdaSpace{}%
\AgdaSymbol{(}\AgdaFunction{A}\AgdaSpace{}%
\AgdaBound{l}\AgdaSymbol{))}\AgdaSpace{}%
\AgdaSymbol{:}\AgdaSpace{}%
\AgdaPrimitiveType{Set}\AgdaSpace{}%
\AgdaBound{a}\AgdaSpace{}%
\AgdaKeyword{where}\<%
\\
\>[0][@{}l@{\AgdaIndent{0}}]%
\>[2]\AgdaInductiveConstructor{imap}\AgdaSpace{}%
\AgdaSymbol{:}\AgdaSpace{}%
\AgdaSymbol{(}\AgdaField{Po}\AgdaSpace{}%
\AgdaSymbol{(}\AgdaFunction{A}\AgdaSpace{}%
\AgdaBound{l}\AgdaSymbol{)}\AgdaSpace{}%
\AgdaBound{s}\AgdaSpace{}%
\AgdaSymbol{→}\AgdaSpace{}%
\AgdaBound{X}\AgdaSymbol{)}\AgdaSpace{}%
\AgdaSymbol{→}\AgdaSpace{}%
\AgdaDatatype{Ar}\AgdaSpace{}%
\AgdaBound{l}\AgdaSpace{}%
\AgdaBound{X}\AgdaSpace{}%
\AgdaBound{s}\<%
\end{code}
Note that instead of fixing \texttt{A} in \texttt{Ar}, we can have a generic
definition:
\begin{code}%
\>[0]\AgdaKeyword{data}\AgdaSpace{}%
\AgdaDatatype{TC}\AgdaSpace{}%
\AgdaSymbol{\{}\AgdaBound{C}\AgdaSpace{}%
\AgdaSymbol{:}\AgdaSpace{}%
\AgdaRecord{Con}\AgdaSymbol{\}\{}\AgdaBound{a}\AgdaSymbol{\}}\AgdaSpace{}%
\AgdaSymbol{(}\AgdaBound{X}\AgdaSpace{}%
\AgdaSymbol{:}\AgdaSpace{}%
\AgdaPrimitiveType{Set}\AgdaSpace{}%
\AgdaBound{a}\AgdaSymbol{)}\AgdaSpace{}%
\AgdaSymbol{(}\AgdaBound{s}\AgdaSpace{}%
\AgdaSymbol{:}\AgdaSpace{}%
\AgdaField{Sh}\AgdaSpace{}%
\AgdaBound{C}\AgdaSymbol{)}\AgdaSpace{}%
\AgdaSymbol{:}\AgdaSpace{}%
\AgdaPrimitiveType{Set}\AgdaSpace{}%
\AgdaBound{a}\AgdaSpace{}%
\AgdaKeyword{where}\<%
\\
\>[0][@{}l@{\AgdaIndent{0}}]%
\>[2]\AgdaInductiveConstructor{imap}\AgdaSpace{}%
\AgdaSymbol{:}\AgdaSpace{}%
\AgdaSymbol{(}\AgdaField{Po}\AgdaSpace{}%
\AgdaBound{C}\AgdaSpace{}%
\AgdaBound{s}\AgdaSpace{}%
\AgdaSymbol{→}\AgdaSpace{}%
\AgdaBound{X}\AgdaSymbol{)}\AgdaSpace{}%
\AgdaSymbol{→}\AgdaSpace{}%
\AgdaDatatype{TC}\AgdaSpace{}%
\AgdaBound{X}\AgdaSpace{}%
\AgdaBound{s}\<%
\end{code}
in which case \texttt{Ar l X s} would be defined as \texttt{TC \{C = A l\} X s}.
In some sense \texttt{C} and \texttt{TC} (or \texttt{A} and \texttt{Ar}) are
related in a similar way as \texttt{List} and \texttt{Vec}.  We lift a commonly
used invariant (shape in case of containers and length in case of lists) into
a type-level argument.

With the help of \texttt{Ar} we can express \texttt{cons} as:
\begin{code}%
\>[0]\AgdaPostulate{cons-ar}\AgdaSpace{}%
\AgdaSymbol{:}\AgdaSpace{}%
\AgdaSymbol{∀}\AgdaSpace{}%
\AgdaSymbol{\{}\AgdaBound{X}\AgdaSpace{}%
\AgdaBound{s}\AgdaSymbol{\}}\AgdaSpace{}%
\AgdaSymbol{→}\AgdaSpace{}%
\AgdaBound{X}\AgdaSpace{}%
\AgdaSymbol{→}\AgdaSpace{}%
\AgdaDatatype{Ar}\AgdaSpace{}%
\AgdaNumber{1}\AgdaSpace{}%
\AgdaBound{X}\AgdaSpace{}%
\AgdaSymbol{(\AgdaUnderscore{}}\AgdaSpace{}%
\AgdaOperator{\AgdaInductiveConstructor{,}}\AgdaSpace{}%
\AgdaBound{s}\AgdaSymbol{)}\AgdaSpace{}%
\AgdaSymbol{→}\AgdaSpace{}%
\AgdaDatatype{Ar}\AgdaSpace{}%
\AgdaNumber{1}\AgdaSpace{}%
\AgdaBound{X}\AgdaSpace{}%
\AgdaSymbol{(\AgdaUnderscore{}}\AgdaSpace{}%
\AgdaOperator{\AgdaInductiveConstructor{,}}\AgdaSpace{}%
\AgdaInductiveConstructor{suc}\AgdaSpace{}%
\AgdaOperator{\AgdaFunction{∘}}\AgdaSpace{}%
\AgdaBound{s}\AgdaSymbol{)}\<%
\end{code}
Also, such a formulation naturally gives rise to the \texttt{imap}
construct that is a basic building block of the SaC programming language.
In SaC arrays are treated as tabulated index-value functions.  The \texttt{imap}
construct can be thought of as an abstract tag that indicates that a chosen
function has to be eventually tabulated.  However, the exact details on how
this function is to be tabulated are not specified.  As a result, when producing
an executable for the given program, a compiler has 
a lot of freedom to choose storage formats for arrays, based on the information
that is being accumulated during optimisation phases.

The two data structures are isomorphic, which is indicated by the following
conversion functions
\begin{code}%
\>[0]\AgdaFunction{c→ar}\AgdaSpace{}%
\AgdaSymbol{:}\AgdaSpace{}%
\AgdaSymbol{∀}\AgdaSpace{}%
\AgdaSymbol{\{}\AgdaBound{n}\AgdaSpace{}%
\AgdaBound{X}\AgdaSymbol{\}}\AgdaSpace{}%
\AgdaSymbol{→}\AgdaSpace{}%
\AgdaSymbol{(}\AgdaBound{c}\AgdaSpace{}%
\AgdaSymbol{:}\AgdaSpace{}%
\AgdaOperator{\AgdaFunction{⟦}}\AgdaSpace{}%
\AgdaFunction{A}\AgdaSpace{}%
\AgdaBound{n}\AgdaSpace{}%
\AgdaOperator{\AgdaFunction{$\rrbracket_{\lhd}$}}\AgdaSpace{}%
\AgdaBound{X}\AgdaSymbol{)}\AgdaSpace{}%
\AgdaSymbol{→}\AgdaSpace{}%
\AgdaDatatype{Ar}\AgdaSpace{}%
\AgdaBound{n}\AgdaSpace{}%
\AgdaBound{X}\AgdaSpace{}%
\AgdaSymbol{(}\AgdaField{proj₁}\AgdaSpace{}%
\AgdaBound{c}\AgdaSymbol{)}\<%
\\
\>[0]\AgdaFunction{c→ar}\AgdaSpace{}%
\AgdaBound{c}\AgdaSpace{}%
\AgdaSymbol{=}\AgdaSpace{}%
\AgdaInductiveConstructor{imap}\AgdaSpace{}%
\AgdaSymbol{(}\AgdaField{proj₂}\AgdaSpace{}%
\AgdaBound{c}\AgdaSymbol{)}\<%
\\
\\[\AgdaEmptyExtraSkip]%
\>[0]\AgdaFunction{ar→c}\AgdaSpace{}%
\AgdaSymbol{:}\AgdaSpace{}%
\AgdaSymbol{∀}\AgdaSpace{}%
\AgdaSymbol{\{}\AgdaBound{n}\AgdaSpace{}%
\AgdaBound{X}\AgdaSpace{}%
\AgdaBound{s}\AgdaSymbol{\}}\AgdaSpace{}%
\AgdaSymbol{→}\AgdaSpace{}%
\AgdaDatatype{Ar}\AgdaSpace{}%
\AgdaBound{n}\AgdaSpace{}%
\AgdaBound{X}\AgdaSpace{}%
\AgdaBound{s}\AgdaSpace{}%
\AgdaSymbol{→}\AgdaSpace{}%
\AgdaOperator{\AgdaFunction{⟦}}\AgdaSpace{}%
\AgdaFunction{A}\AgdaSpace{}%
\AgdaBound{n}\AgdaSpace{}%
\AgdaOperator{\AgdaFunction{$\rrbracket_{\lhd}$}}\AgdaSpace{}%
\AgdaBound{X}\<%
\\
\>[0]\AgdaFunction{ar→c}\AgdaSpace{}%
\AgdaSymbol{\{}\AgdaArgument{s}\AgdaSpace{}%
\AgdaSymbol{=}\AgdaSpace{}%
\AgdaBound{s}\AgdaSymbol{\}}\AgdaSpace{}%
\AgdaSymbol{(}\AgdaInductiveConstructor{imap}\AgdaSpace{}%
\AgdaBound{x}\AgdaSymbol{)}\AgdaSpace{}%
\AgdaSymbol{=}\AgdaSpace{}%
\AgdaBound{s}\AgdaSpace{}%
\AgdaOperator{\AgdaInductiveConstructor{,}}\AgdaSpace{}%
\AgdaBound{x}\<%
\end{code}

Finally, if we consider \texttt{imap} as an array constructor, there has
to be an eliminator.  The eliminator for the array is a selection operation,
and our 
model, this is simply a function application.
\begin{code}%
\>[0]\AgdaFunction{sel}\AgdaSpace{}%
\AgdaSymbol{:}\AgdaSpace{}%
\AgdaSymbol{∀}\AgdaSpace{}%
\AgdaSymbol{\{}\AgdaBound{a}\AgdaSymbol{\}\{}\AgdaBound{X}\AgdaSpace{}%
\AgdaSymbol{:}\AgdaSpace{}%
\AgdaPrimitiveType{Set}\AgdaSpace{}%
\AgdaBound{a}\AgdaSymbol{\}\{}\AgdaBound{n}\AgdaSpace{}%
\AgdaBound{s}\AgdaSymbol{\}}\AgdaSpace{}%
\AgdaSymbol{→}\AgdaSpace{}%
\AgdaDatatype{Ar}\AgdaSpace{}%
\AgdaBound{n}\AgdaSpace{}%
\AgdaBound{X}\AgdaSpace{}%
\AgdaBound{s}\AgdaSpace{}%
\AgdaSymbol{→}\AgdaSpace{}%
\AgdaField{Po}\AgdaSpace{}%
\AgdaSymbol{(}\AgdaFunction{A}\AgdaSpace{}%
\AgdaBound{n}\AgdaSymbol{)}\AgdaSpace{}%
\AgdaBound{s}\AgdaSpace{}%
\AgdaSymbol{→}\AgdaSpace{}%
\AgdaBound{X}\<%
\\
\>[0]\AgdaFunction{sel}\AgdaSpace{}%
\AgdaSymbol{(}\AgdaInductiveConstructor{imap}\AgdaSpace{}%
\AgdaBound{x}\AgdaSymbol{)}\AgdaSpace{}%
\AgdaBound{iv}\AgdaSpace{}%
\AgdaSymbol{=}\AgdaSpace{}%
\AgdaBound{x}\AgdaSpace{}%
\AgdaBound{iv}\<%
\end{code}

\subsection{Lack of Extensionality}

While the above model gives us all the fundamental array primitives, it
has a serious flaw when we come to reasoning about array equalities.
One of the fundamental assumptions in array calculi is that the same
indices select the same values.  Indices of level-$n$ arrays, where
$n > 0$ are functions, and therefore we define index equality extensionally.
It should not matter how exactly the elements within the index are computed,
as long as two indices are element-wise equal, they should select the same
element.  Unfortunately, in Agda this cannot be shown.  The code below
demonstrates the problem.

\begin{code}%
\>[0]\AgdaFunction{po-eq}\AgdaSpace{}%
\AgdaSymbol{:}\AgdaSpace{}%
\AgdaSymbol{∀}\AgdaSpace{}%
\AgdaSymbol{\{}\AgdaBound{l}\AgdaSpace{}%
\AgdaBound{s}\AgdaSymbol{\}}\AgdaSpace{}%
\AgdaSymbol{→}\AgdaSpace{}%
\AgdaSymbol{(}\AgdaBound{iv}\AgdaSpace{}%
\AgdaBound{jv}\AgdaSpace{}%
\AgdaSymbol{:}\AgdaSpace{}%
\AgdaField{Po}\AgdaSpace{}%
\AgdaSymbol{(}\AgdaFunction{A}\AgdaSpace{}%
\AgdaBound{l}\AgdaSymbol{)}\AgdaSpace{}%
\AgdaBound{s}\AgdaSymbol{)}\AgdaSpace{}%
\AgdaSymbol{→}\AgdaSpace{}%
\AgdaPrimitiveType{Set}\<%
\\
\>[0]\AgdaFunction{po-eq}\AgdaSpace{}%
\AgdaSymbol{\{}\AgdaInductiveConstructor{zero}\AgdaSymbol{\}}%
\>[14]\AgdaBound{iv}\AgdaSpace{}%
\AgdaBound{jv}\AgdaSpace{}%
\AgdaSymbol{=}\AgdaSpace{}%
\AgdaBound{iv}\AgdaSpace{}%
\AgdaOperator{\AgdaDatatype{≡}}\AgdaSpace{}%
\AgdaBound{jv}\<%
\\
\>[0]\AgdaFunction{po-eq}\AgdaSpace{}%
\AgdaSymbol{\{}\AgdaInductiveConstructor{suc}\AgdaSpace{}%
\AgdaBound{l}\AgdaSymbol{\}}\AgdaSpace{}%
\AgdaBound{iv}\AgdaSpace{}%
\AgdaBound{jv}\AgdaSpace{}%
\AgdaSymbol{=}\AgdaSpace{}%
\AgdaSymbol{∀}\AgdaSpace{}%
\AgdaBound{i}\AgdaSpace{}%
\AgdaSymbol{→}\AgdaSpace{}%
\AgdaBound{iv}\AgdaSpace{}%
\AgdaBound{i}\AgdaSpace{}%
\AgdaOperator{\AgdaDatatype{≡}}\AgdaSpace{}%
\AgdaBound{jv}\AgdaSpace{}%
\AgdaBound{i}\<%
\\
\\[\AgdaEmptyExtraSkip]%
\>[0]\AgdaPostulate{sel-eq}%
\>[384I]\AgdaSymbol{:}\AgdaSpace{}%
\AgdaSymbol{∀}\AgdaSpace{}%
\AgdaSymbol{\{}\AgdaBound{a}\AgdaSymbol{\}\{}\AgdaBound{X}\AgdaSpace{}%
\AgdaSymbol{:}\AgdaSpace{}%
\AgdaPrimitiveType{Set}\AgdaSpace{}%
\AgdaBound{a}\AgdaSymbol{\}\{}\AgdaBound{l}\AgdaSpace{}%
\AgdaBound{s}\AgdaSymbol{\}}\<%
\\
\>[.][@{}l@{}]\<[384I]%
\>[7]\AgdaSymbol{→}\AgdaSpace{}%
\AgdaSymbol{(}\AgdaBound{a}\AgdaSpace{}%
\AgdaSymbol{:}\AgdaSpace{}%
\AgdaDatatype{Ar}\AgdaSpace{}%
\AgdaBound{l}\AgdaSpace{}%
\AgdaBound{X}\AgdaSpace{}%
\AgdaBound{s}\AgdaSymbol{)}\<%
\\
\>[7]\AgdaSymbol{→}\AgdaSpace{}%
\AgdaSymbol{(}\AgdaBound{iv}\AgdaSpace{}%
\AgdaBound{jv}\AgdaSpace{}%
\AgdaSymbol{:}\AgdaSpace{}%
\AgdaField{Po}\AgdaSpace{}%
\AgdaSymbol{(}\AgdaFunction{A}\AgdaSpace{}%
\AgdaBound{l}\AgdaSymbol{)}\AgdaSpace{}%
\AgdaBound{s}\AgdaSymbol{)}\<%
\\
\>[7]\AgdaSymbol{→}\AgdaSpace{}%
\AgdaFunction{po-eq}\AgdaSpace{}%
\AgdaSymbol{\{}\AgdaArgument{l}\AgdaSpace{}%
\AgdaSymbol{=}\AgdaSpace{}%
\AgdaBound{l}\AgdaSymbol{\}}\AgdaSpace{}%
\AgdaBound{iv}\AgdaSpace{}%
\AgdaBound{jv}\<%
\\
\>[7]\AgdaSymbol{→}\AgdaSpace{}%
\AgdaFunction{sel}\AgdaSpace{}%
\AgdaBound{a}\AgdaSpace{}%
\AgdaBound{iv}\AgdaSpace{}%
\AgdaOperator{\AgdaDatatype{≡}}\AgdaSpace{}%
\AgdaFunction{sel}\AgdaSpace{}%
\AgdaBound{a}\AgdaSpace{}%
\AgdaBound{jv}\<%
\end{code}
Recall that even at the first level the indices have a type \texttt{⊤ → Fin (s tt)}:
\begin{code}%
\>[0]\AgdaFunction{sanity}\AgdaSpace{}%
\AgdaSymbol{:}\AgdaSpace{}%
\AgdaSymbol{∀}\AgdaSpace{}%
\AgdaBound{s}\AgdaSpace{}%
\AgdaSymbol{→}\AgdaSpace{}%
\AgdaField{Po}\AgdaSpace{}%
\AgdaSymbol{(}\AgdaFunction{A}\AgdaSpace{}%
\AgdaNumber{1}\AgdaSymbol{)}\AgdaSpace{}%
\AgdaSymbol{(\AgdaUnderscore{}}\AgdaSpace{}%
\AgdaOperator{\AgdaInductiveConstructor{,}}\AgdaSpace{}%
\AgdaBound{s}\AgdaSymbol{)}\AgdaSpace{}%
\AgdaOperator{\AgdaDatatype{≡}}\AgdaSpace{}%
\AgdaSymbol{(}\AgdaRecord{⊤}\AgdaSpace{}%
\AgdaSymbol{→}\AgdaSpace{}%
\AgdaDatatype{Fin}\AgdaSpace{}%
\AgdaSymbol{(}\AgdaBound{s}\AgdaSpace{}%
\AgdaInductiveConstructor{tt}\AgdaSymbol{))}\<%
\\
\>[0]\AgdaFunction{sanity}\AgdaSpace{}%
\AgdaBound{s}\AgdaSpace{}%
\AgdaSymbol{=}\AgdaSpace{}%
\AgdaInductiveConstructor{refl}\<%
\end{code}

Without the \texttt{sel-eq} property we cannot show very fundamental array facts such as
element preservation under flattening/unflattening, reshaping, transposition,
\etc{} This problem can be worked around in a number of ways including defining
custom equality relation and working in setoids, defining custom selection operation
or using cubical Agda.
In this paper we introduces a non-functional representation for indices so that the
\texttt{sel-eq} becomes provable.  The details are described in the next sections.

 \section{Alternative Encoding}
\begin{code}[hide]%
\>[0]\AgdaSymbol{\{-\#}\AgdaSpace{}%
\AgdaKeyword{OPTIONS}\AgdaSpace{}%
\AgdaPragma{--rewriting}\AgdaSpace{}%
\AgdaPragma{--inversion-max-depth=100}%
\>[51]\AgdaSymbol{\#-\}}\<%
\\
\>[0]\AgdaKeyword{open}\AgdaSpace{}%
\AgdaKeyword{import}\AgdaSpace{}%
\AgdaModule{Data.Nat}\<%
\\
\>[0]\AgdaKeyword{open}\AgdaSpace{}%
\AgdaKeyword{import}\AgdaSpace{}%
\AgdaModule{Data.Nat.DivMod}\<%
\\
\>[0]\AgdaKeyword{open}\AgdaSpace{}%
\AgdaKeyword{import}\AgdaSpace{}%
\AgdaModule{Data.Nat.Properties}\<%
\\
\\[\AgdaEmptyExtraSkip]%
\>[0]\AgdaKeyword{open}\AgdaSpace{}%
\AgdaKeyword{import}\AgdaSpace{}%
\AgdaModule{Data.Product}\<%
\\
\>[0]\AgdaKeyword{open}\AgdaSpace{}%
\AgdaKeyword{import}\AgdaSpace{}%
\AgdaModule{Data.Sum}\<%
\\
\>[0]\AgdaKeyword{open}\AgdaSpace{}%
\AgdaKeyword{import}\AgdaSpace{}%
\AgdaModule{Function}\AgdaSpace{}%
\AgdaKeyword{using}\AgdaSpace{}%
\AgdaSymbol{(}\AgdaOperator{\AgdaFunction{\AgdaUnderscore{}\$\AgdaUnderscore{}}}\AgdaSpace{}%
\AgdaSymbol{;}\AgdaSpace{}%
\AgdaOperator{\AgdaFunction{\AgdaUnderscore{}∘\AgdaUnderscore{}}}\AgdaSpace{}%
\AgdaSymbol{;}\AgdaSpace{}%
\AgdaOperator{\AgdaFunction{case\AgdaUnderscore{}of\AgdaUnderscore{}}}\AgdaSymbol{)}\<%
\\
\>[0]\AgdaKeyword{open}\AgdaSpace{}%
\AgdaKeyword{import}\AgdaSpace{}%
\AgdaModule{Data.Vec}\AgdaSpace{}%
\AgdaKeyword{hiding}\AgdaSpace{}%
\AgdaSymbol{(}\AgdaFunction{sum}\AgdaSymbol{)}\<%
\\
\>[0]\AgdaKeyword{open}\AgdaSpace{}%
\AgdaKeyword{import}\AgdaSpace{}%
\AgdaModule{Data.Vec.Properties}\<%
\\
\>[0]\AgdaKeyword{open}\AgdaSpace{}%
\AgdaKeyword{import}\AgdaSpace{}%
\AgdaModule{Data.Unit}\AgdaSpace{}%
\AgdaKeyword{hiding}\AgdaSpace{}%
\AgdaSymbol{(}\AgdaOperator{\AgdaFunction{\AgdaUnderscore{}≟\AgdaUnderscore{}}}\AgdaSymbol{;}\AgdaSpace{}%
\AgdaOperator{\AgdaRecord{\AgdaUnderscore{}≤\AgdaUnderscore{}}}\AgdaSymbol{)}\<%
\\
\>[0]\AgdaComment{\ {-}{-}open import Level hiding (suc)}\<%
\\
\>[0]\<%
\end{code}
The main idea here is to define an alternative representation for \texttt{A}
from the previous section that makes it possible to prove that indices with the
same components select the same array elements.  We achieve this by using a
non-function based representation for shapes and indices, while still representing
contents of arrays as imap-tagged index-value functions.

We start with an alternative definition for \AgdaDatatype{Fin} in a refinement
type~\cite{refinement-ml,refinement-logic} style.  While this is
not strictly necessary, it helps to simplify a number of proofs later.  The main
difference from the regular \AgdaDatatype{Fin} is that we keep the actual value
of an index as an element of type \AgdaDatatype{ℕ}, and make irrelevant (in Agda sense) the proof that this
value is less than the chosen upper bound.  Note the dot in the \texttt{v<u} field
name.

\begin{code}%
\>[0]\AgdaKeyword{record}\AgdaSpace{}%
\AgdaRecord{BFin}\AgdaSpace{}%
\AgdaSymbol{(}\AgdaBound{u}\AgdaSpace{}%
\AgdaSymbol{:}\AgdaSpace{}%
\AgdaDatatype{ℕ}\AgdaSymbol{)}\AgdaSpace{}%
\AgdaSymbol{:}\AgdaSpace{}%
\AgdaPrimitiveType{Set}\AgdaSpace{}%
\AgdaKeyword{where}\<%
\\
\>[0][@{}l@{\AgdaIndent{0}}]%
\>[2]\AgdaKeyword{constructor}\AgdaSpace{}%
\AgdaOperator{\AgdaInductiveConstructor{\AgdaUnderscore{}bounded\AgdaUnderscore{}}}\<%
\\
\>[2]\AgdaKeyword{field}\<%
\\
\>[2][@{}l@{\AgdaIndent{0}}]%
\>[4]\AgdaField{v}\AgdaSpace{}%
\AgdaSymbol{:}\AgdaSpace{}%
\AgdaDatatype{ℕ}\<%
\\
\>[4]\AgdaSymbol{.}\AgdaField{v<u}\AgdaSpace{}%
\AgdaSymbol{:}\AgdaSpace{}%
\AgdaBound{v}\AgdaSpace{}%
\AgdaOperator{\AgdaFunction{<}}\AgdaSpace{}%
\AgdaBound{u}\<%
\end{code}
\begin{code}[hide]%
\>[0]\AgdaKeyword{open}\AgdaSpace{}%
\AgdaModule{BFin}\<%
\end{code}

As we are defining an array and its representation at the same time,
and we make these definitions interdependent, we sometimes have to provide
types first and only later provide the actual definition.  We start with
leveled types for shapes, indices and array representations.
\begin{code}%
\>[0]\AgdaFunction{ShType}\AgdaSpace{}%
\AgdaSymbol{:}\AgdaSpace{}%
\AgdaSymbol{(}\AgdaBound{l}\AgdaSpace{}%
\AgdaSymbol{:}\AgdaSpace{}%
\AgdaDatatype{ℕ}\AgdaSymbol{)}\AgdaSpace{}%
\AgdaSymbol{→}\AgdaSpace{}%
\AgdaPrimitiveType{Set}\<%
\\
\>[0]\AgdaFunction{IxType}\AgdaSpace{}%
\AgdaSymbol{:}\AgdaSpace{}%
\AgdaSymbol{(}\AgdaBound{l}\AgdaSpace{}%
\AgdaSymbol{:}\AgdaSpace{}%
\AgdaDatatype{ℕ}\AgdaSymbol{)}\AgdaSpace{}%
\AgdaSymbol{→}\AgdaSpace{}%
\AgdaFunction{ShType}\AgdaSpace{}%
\AgdaBound{l}\AgdaSpace{}%
\AgdaSymbol{→}\AgdaSpace{}%
\AgdaPrimitiveType{Set}\<%
\\
\>[0]\AgdaFunction{ReprAr}\AgdaSpace{}%
\AgdaSymbol{:}\AgdaSpace{}%
\AgdaSymbol{(}\AgdaBound{l}\AgdaSpace{}%
\AgdaSymbol{:}\AgdaSpace{}%
\AgdaDatatype{ℕ}\AgdaSymbol{)}\AgdaSpace{}%
\AgdaSymbol{→}\AgdaSpace{}%
\AgdaSymbol{(}\AgdaBound{X}\AgdaSpace{}%
\AgdaSymbol{:}\AgdaSpace{}%
\AgdaPrimitiveType{Set}\AgdaSymbol{)}\AgdaSpace{}%
\AgdaSymbol{→}\AgdaSpace{}%
\AgdaPrimitiveType{Set}\<%
\end{code}
Note that for simplicity we do not impose any requirements on the
validity of array representation.  We can define these properties later
extrinsically.

As will be seen later, \AgdaDatatype{IxType}
may have the same representation for indices into arrays of different shapes.
To avoid this we define the \AgdaDatatype{Ix} type that
wraps \AgdaDatatype{IxType} and carries array shape as a type parameter.
\begin{code}%
\>[0]\AgdaKeyword{record}\AgdaSpace{}%
\AgdaRecord{Ix}\AgdaSpace{}%
\AgdaSymbol{(}\AgdaBound{l}\AgdaSpace{}%
\AgdaSymbol{:}\AgdaSpace{}%
\AgdaDatatype{ℕ}\AgdaSymbol{)}\AgdaSpace{}%
\AgdaSymbol{(}\AgdaBound{s}\AgdaSpace{}%
\AgdaSymbol{:}\AgdaSpace{}%
\AgdaFunction{ShType}\AgdaSpace{}%
\AgdaBound{l}\AgdaSymbol{)}\AgdaSpace{}%
\AgdaSymbol{:}\AgdaSpace{}%
\AgdaPrimitiveType{Set}\AgdaSpace{}%
\AgdaKeyword{where}\<%
\\
\>[0][@{}l@{\AgdaIndent{0}}]%
\>[2]\AgdaKeyword{constructor}\AgdaSpace{}%
\AgdaInductiveConstructor{ix}\<%
\\
\>[2]\AgdaKeyword{field}\<%
\\
\>[2][@{}l@{\AgdaIndent{0}}]%
\>[4]\AgdaField{flat-ix}\AgdaSpace{}%
\AgdaSymbol{:}\AgdaSpace{}%
\AgdaFunction{IxType}\AgdaSpace{}%
\AgdaBound{l}\AgdaSpace{}%
\AgdaBound{s}\<%
\end{code}
As a result, we will not be able to index an array of shape $s$ with an
index bound by shape $s_1$ without explicit cast.  In the same way as one
cannot pass the term of type \texttt{Fin\ 10} to the function \texttt{Fin\ 15 → X}.

As before, arrays are tabulated index-value functions where \texttt{imap}
is a constructor.
\begin{code}%
\>[0]\AgdaKeyword{data}\AgdaSpace{}%
\AgdaDatatype{Ar}\AgdaSpace{}%
\AgdaSymbol{\{}\AgdaBound{a}\AgdaSymbol{\}}\AgdaSpace{}%
\AgdaSymbol{(}\AgdaBound{l}\AgdaSpace{}%
\AgdaSymbol{:}\AgdaSpace{}%
\AgdaDatatype{ℕ}\AgdaSymbol{)}\AgdaSpace{}%
\AgdaSymbol{(}\AgdaBound{X}\AgdaSpace{}%
\AgdaSymbol{:}\AgdaSpace{}%
\AgdaPrimitiveType{Set}\AgdaSymbol{)}\AgdaSpace{}%
\AgdaSymbol{(}\AgdaBound{s}\AgdaSpace{}%
\AgdaSymbol{:}\AgdaSpace{}%
\AgdaFunction{ShType}\AgdaSpace{}%
\AgdaBound{l}\AgdaSymbol{)}\AgdaSpace{}%
\AgdaSymbol{:}\AgdaSpace{}%
\AgdaPrimitiveType{Set}\AgdaSpace{}%
\AgdaBound{a}\AgdaSpace{}%
\AgdaKeyword{where}\<%
\\
\>[0][@{}l@{\AgdaIndent{0}}]%
\>[2]\AgdaInductiveConstructor{imap}\AgdaSpace{}%
\AgdaSymbol{:}\AgdaSpace{}%
\AgdaSymbol{(}\AgdaRecord{Ix}\AgdaSpace{}%
\AgdaBound{l}\AgdaSpace{}%
\AgdaBound{s}\AgdaSpace{}%
\AgdaSymbol{→}\AgdaSpace{}%
\AgdaBound{X}\AgdaSymbol{)}\AgdaSpace{}%
\AgdaSymbol{→}\AgdaSpace{}%
\AgdaDatatype{Ar}\AgdaSpace{}%
\AgdaBound{l}\AgdaSpace{}%
\AgdaBound{X}\AgdaSpace{}%
\AgdaBound{s}\<%
\end{code}
Our array representation needs the information on how many elements
does the array contain, which is just a product of the shape elements.
Again, we cannot yet provide the body of the function, as we have not
yet defined how we represent shapes.
\begin{code}%
\>[0]\AgdaFunction{prod}\AgdaSpace{}%
\AgdaSymbol{:}\AgdaSpace{}%
\AgdaSymbol{∀}\AgdaSpace{}%
\AgdaSymbol{\{}\AgdaBound{l}\AgdaSymbol{\}}\AgdaSpace{}%
\AgdaSymbol{→}\AgdaSpace{}%
\AgdaFunction{ShType}\AgdaSpace{}%
\AgdaBound{l}\AgdaSpace{}%
\AgdaSymbol{→}\AgdaSpace{}%
\AgdaDatatype{ℕ}\<%
\end{code}

Finally, we get to the definition of the shape representation, which
we chose to be the unit type for level-0 arrays and \AgdaDatatype{ReprAr}
otherwise.  While the latter is not strictly necessary, it gives us
a nice symmetry between the arrays and their shapes.
\begin{code}%
\>[0]\AgdaFunction{ShType}\AgdaSpace{}%
\AgdaInductiveConstructor{zero}%
\>[15]\AgdaSymbol{=}\AgdaSpace{}%
\AgdaRecord{⊤}\<%
\\
\>[0]\AgdaFunction{ShType}\AgdaSpace{}%
\AgdaSymbol{(}\AgdaInductiveConstructor{suc}\AgdaSpace{}%
\AgdaBound{l}\AgdaSymbol{)}\AgdaSpace{}%
\AgdaSymbol{=}\AgdaSpace{}%
\AgdaFunction{ReprAr}\AgdaSpace{}%
\AgdaBound{l}\AgdaSpace{}%
\AgdaDatatype{ℕ}\<%
\end{code}

Our array representation is a dependent pair where the first element
is the representation of the shape, and the second element is a
linearisation of array elements --- a vector that has as many elements
as an array of the given shape.
\begin{code}%
\>[0]\AgdaFunction{ReprAr}\AgdaSpace{}%
\AgdaBound{l}\AgdaSpace{}%
\AgdaBound{X}\AgdaSpace{}%
\AgdaSymbol{=}\AgdaSpace{}%
\AgdaRecord{Σ}\AgdaSpace{}%
\AgdaSymbol{(}\AgdaFunction{ShType}\AgdaSpace{}%
\AgdaBound{l}\AgdaSymbol{)}\AgdaSpace{}%
\AgdaSymbol{λ}\AgdaSpace{}%
\AgdaBound{s}\AgdaSpace{}%
\AgdaSymbol{→}\AgdaSpace{}%
\AgdaDatatype{Vec}\AgdaSpace{}%
\AgdaBound{X}\AgdaSpace{}%
\AgdaSymbol{(}\AgdaFunction{prod}\AgdaSpace{}%
\AgdaSymbol{\{}\AgdaArgument{l}\AgdaSpace{}%
\AgdaSymbol{=}\AgdaSpace{}%
\AgdaBound{l}\AgdaSymbol{\}}\AgdaSpace{}%
\AgdaBound{s}\AgdaSymbol{)}\<%
\end{code}

As before, indices into an array of shape $s$ have the same structure
as $s$, except each valid index is component-wise less than $s$.
As \AgdaDatatype{ShType} keeps shape elements in a linearised form,
the index type mimics the structure of the shape:
\begin{code}%
\>[0]\AgdaKeyword{infixr}\AgdaSpace{}%
\AgdaNumber{5}\AgdaSpace{}%
\AgdaOperator{\AgdaInductiveConstructor{\AgdaUnderscore{}∷\AgdaUnderscore{}}}\<%
\\
\>[0]\AgdaKeyword{data}\AgdaSpace{}%
\AgdaDatatype{FlatIx}\AgdaSpace{}%
\AgdaSymbol{:}\AgdaSpace{}%
\AgdaSymbol{(}\AgdaBound{d}\AgdaSpace{}%
\AgdaSymbol{:}\AgdaSpace{}%
\AgdaDatatype{ℕ}\AgdaSymbol{)}\AgdaSpace{}%
\AgdaSymbol{→}\AgdaSpace{}%
\AgdaSymbol{(}\AgdaBound{s}\AgdaSpace{}%
\AgdaSymbol{:}\AgdaSpace{}%
\AgdaDatatype{Vec}\AgdaSpace{}%
\AgdaDatatype{ℕ}\AgdaSpace{}%
\AgdaBound{d}\AgdaSymbol{)}\AgdaSpace{}%
\AgdaSymbol{→}\AgdaSpace{}%
\AgdaPrimitiveType{Set}\AgdaSpace{}%
\AgdaKeyword{where}\<%
\\
\>[0][@{}l@{\AgdaIndent{0}}]%
\>[2]\AgdaInductiveConstructor{[]}%
\>[7]\AgdaSymbol{:}\AgdaSpace{}%
\AgdaDatatype{FlatIx}\AgdaSpace{}%
\AgdaNumber{0}\AgdaSpace{}%
\AgdaInductiveConstructor{[]}\<%
\\
\>[2]\AgdaOperator{\AgdaInductiveConstructor{\AgdaUnderscore{}∷\AgdaUnderscore{}}}\AgdaSpace{}%
\AgdaSymbol{:}\AgdaSpace{}%
\AgdaSymbol{∀}\AgdaSpace{}%
\AgdaSymbol{\{}\AgdaBound{d}\AgdaSpace{}%
\AgdaBound{s}\AgdaSpace{}%
\AgdaBound{x}\AgdaSymbol{\}}\AgdaSpace{}%
\AgdaSymbol{→}\AgdaSpace{}%
\AgdaRecord{BFin}\AgdaSpace{}%
\AgdaBound{x}\AgdaSpace{}%
\AgdaSymbol{→}\AgdaSpace{}%
\AgdaSymbol{(}\AgdaBound{ix}\AgdaSpace{}%
\AgdaSymbol{:}\AgdaSpace{}%
\AgdaDatatype{FlatIx}\AgdaSpace{}%
\AgdaBound{d}\AgdaSpace{}%
\AgdaBound{s}\AgdaSymbol{)}\AgdaSpace{}%
\AgdaSymbol{→}\AgdaSpace{}%
\AgdaDatatype{FlatIx}\AgdaSpace{}%
\AgdaSymbol{(}\AgdaInductiveConstructor{suc}\AgdaSpace{}%
\AgdaBound{d}\AgdaSymbol{)}\AgdaSpace{}%
\AgdaSymbol{(}\AgdaBound{x}\AgdaSpace{}%
\AgdaOperator{\AgdaInductiveConstructor{∷}}\AgdaSpace{}%
\AgdaBound{s}\AgdaSymbol{)}\<%
\end{code}
That is, a linearised shape is given by a vector of $d$ elements,
therefore an index contains \AgdaDatatype{BFin} elements where the upper
bounds refer to the corresponding elements of the linearised shape.

The only valid index type for level-0 shapes is a singleton type,
for which we use the unit type, and for higher level arrays we use
\AgdaDatatype{FlatIx}.


\begin{code}%
\>[0]\AgdaFunction{IxType}\AgdaSpace{}%
\AgdaInductiveConstructor{zero}\AgdaSpace{}%
\AgdaInductiveConstructor{tt}\AgdaSpace{}%
\AgdaSymbol{=}\AgdaSpace{}%
\AgdaRecord{⊤}\<%
\\
\>[0]\AgdaFunction{IxType}\AgdaSpace{}%
\AgdaSymbol{(}\AgdaInductiveConstructor{suc}\AgdaSpace{}%
\AgdaBound{l}\AgdaSymbol{)}\AgdaSpace{}%
\AgdaSymbol{(}\AgdaBound{s}\AgdaSpace{}%
\AgdaOperator{\AgdaInductiveConstructor{,}}\AgdaSpace{}%
\AgdaBound{v}\AgdaSymbol{)}\AgdaSpace{}%
\AgdaSymbol{=}\AgdaSpace{}%
\AgdaDatatype{FlatIx}\AgdaSpace{}%
\AgdaSymbol{(}\AgdaFunction{prod}\AgdaSpace{}%
\AgdaBound{s}\AgdaSymbol{)}\AgdaSpace{}%
\AgdaBound{v}\<%
\end{code}

The product is a fold of multiplication with the neutral element 1.
\begin{code}%
\>[0]\AgdaFunction{flat-prod}\AgdaSpace{}%
\AgdaSymbol{:}\AgdaSpace{}%
\AgdaSymbol{∀}\AgdaSpace{}%
\AgdaSymbol{\{}\AgdaBound{n}\AgdaSymbol{\}}\AgdaSpace{}%
\AgdaSymbol{→}\AgdaSpace{}%
\AgdaDatatype{Vec}\AgdaSpace{}%
\AgdaDatatype{ℕ}\AgdaSpace{}%
\AgdaBound{n}\AgdaSpace{}%
\AgdaSymbol{→}\AgdaSpace{}%
\AgdaDatatype{ℕ}\<%
\\
\>[0]\AgdaFunction{flat-prod}\AgdaSpace{}%
\AgdaSymbol{=}\AgdaSpace{}%
\AgdaFunction{foldr}\AgdaSpace{}%
\AgdaSymbol{\AgdaUnderscore{}}\AgdaSpace{}%
\AgdaOperator{\AgdaPrimitive{\AgdaUnderscore{}*\AgdaUnderscore{}}}\AgdaSpace{}%
\AgdaNumber{1}\<%
\\
\\[\AgdaEmptyExtraSkip]%
\>[0]\AgdaFunction{prod}\AgdaSpace{}%
\AgdaSymbol{\{}\AgdaInductiveConstructor{zero}\AgdaSymbol{\}}%
\>[13]\AgdaBound{sh}\AgdaSpace{}%
\AgdaSymbol{=}\AgdaSpace{}%
\AgdaNumber{1}\<%
\\
\>[0]\AgdaFunction{prod}\AgdaSpace{}%
\AgdaSymbol{\{}\AgdaInductiveConstructor{suc}\AgdaSpace{}%
\AgdaBound{l}\AgdaSymbol{\}}\AgdaSpace{}%
\AgdaSymbol{(}\AgdaBound{s}\AgdaSpace{}%
\AgdaOperator{\AgdaInductiveConstructor{,}}\AgdaSpace{}%
\AgdaBound{v}\AgdaSymbol{)}\AgdaSpace{}%
\AgdaSymbol{=}\AgdaSpace{}%
\AgdaFunction{flat-prod}\AgdaSpace{}%
\AgdaBound{v}\<%
\end{code}

\subsection{Examples}
In order to develop a better intuition of the above data structures, let
us define a few simple examples.  We start with defining
some arrays of levels zero, one and two.

\begin{code}%
\>[0]\AgdaFunction{sca}\AgdaSpace{}%
\AgdaSymbol{:}\AgdaSpace{}%
\AgdaDatatype{Ar}\AgdaSpace{}%
\AgdaNumber{0}\AgdaSpace{}%
\AgdaDatatype{ℕ}\AgdaSpace{}%
\AgdaInductiveConstructor{tt}%
\>[44]\AgdaComment{\ {-}{-} A scalar}\<%
\\
\>[0]\AgdaFunction{vec}\AgdaSpace{}%
\AgdaSymbol{:}\AgdaSpace{}%
\AgdaDatatype{Ar}\AgdaSpace{}%
\AgdaNumber{1}\AgdaSpace{}%
\AgdaDatatype{ℕ}\AgdaSpace{}%
\AgdaSymbol{(}\AgdaInductiveConstructor{tt}\AgdaSpace{}%
\AgdaOperator{\AgdaInductiveConstructor{,}}\AgdaSpace{}%
\AgdaNumber{5}\AgdaSpace{}%
\AgdaOperator{\AgdaInductiveConstructor{∷}}\AgdaSpace{}%
\AgdaInductiveConstructor{[]}\AgdaSymbol{)}%
\>[44]\AgdaComment{\ {-}{-} Vector of 5 elements}\<%
\\
\>[0]\AgdaFunction{mat}\AgdaSpace{}%
\AgdaSymbol{:}\AgdaSpace{}%
\AgdaDatatype{Ar}\AgdaSpace{}%
\AgdaNumber{2}\AgdaSpace{}%
\AgdaDatatype{ℕ}\AgdaSpace{}%
\AgdaSymbol{((}\AgdaInductiveConstructor{tt}\AgdaSpace{}%
\AgdaOperator{\AgdaInductiveConstructor{,}}\AgdaSpace{}%
\AgdaNumber{2}\AgdaSpace{}%
\AgdaOperator{\AgdaInductiveConstructor{∷}}\AgdaSpace{}%
\AgdaInductiveConstructor{[]}\AgdaSymbol{)}\AgdaSpace{}%
\AgdaOperator{\AgdaInductiveConstructor{,}}\AgdaSpace{}%
\AgdaNumber{2}\AgdaSpace{}%
\AgdaOperator{\AgdaInductiveConstructor{∷}}\AgdaSpace{}%
\AgdaNumber{2}\AgdaSpace{}%
\AgdaOperator{\AgdaInductiveConstructor{∷}}\AgdaSpace{}%
\AgdaInductiveConstructor{[]}\AgdaSymbol{)}%
\>[44]\AgdaComment{\ {-}{-} Matrix of 2×2 elements}\<%
\end{code}

Let us now define the values. 
\begin{code}%
\>[0]\AgdaFunction{sca}\AgdaSpace{}%
\AgdaSymbol{=}\AgdaSpace{}%
\AgdaInductiveConstructor{imap}\AgdaSpace{}%
\AgdaSymbol{λ}\AgdaSpace{}%
\AgdaBound{\AgdaUnderscore{}}\AgdaSpace{}%
\AgdaSymbol{→}\AgdaSpace{}%
\AgdaNumber{42}\<%
\\
\>[0]\AgdaFunction{vec}\AgdaSpace{}%
\AgdaSymbol{=}\AgdaSpace{}%
\AgdaInductiveConstructor{imap}\AgdaSpace{}%
\AgdaSymbol{λ}\AgdaSpace{}%
\AgdaSymbol{\{(}\AgdaInductiveConstructor{ix}\AgdaSpace{}%
\AgdaSymbol{(}\AgdaNumber{0}\AgdaSpace{}%
\AgdaOperator{\AgdaInductiveConstructor{bounded}}\AgdaSpace{}%
\AgdaSymbol{\AgdaUnderscore{}}\AgdaSpace{}%
\AgdaOperator{\AgdaInductiveConstructor{∷}}\AgdaSpace{}%
\AgdaInductiveConstructor{[]}\AgdaSymbol{))}\AgdaSpace{}%
\AgdaSymbol{→}\AgdaSpace{}%
\AgdaNumber{42}\AgdaSpace{}%
\AgdaSymbol{;}\AgdaSpace{}%
\AgdaSymbol{\AgdaUnderscore{}}\AgdaSpace{}%
\AgdaSymbol{→}\AgdaSpace{}%
\AgdaNumber{0}\AgdaSymbol{\}}\<%
\\
\>[0]\AgdaFunction{mat}\AgdaSpace{}%
\AgdaSymbol{=}\AgdaSpace{}%
\AgdaInductiveConstructor{imap}\AgdaSpace{}%
\AgdaSymbol{λ}\AgdaSpace{}%
\AgdaSymbol{\{(}\AgdaInductiveConstructor{ix}\AgdaSpace{}%
\AgdaSymbol{(}\AgdaNumber{1}\AgdaSpace{}%
\AgdaOperator{\AgdaInductiveConstructor{bounded}}\AgdaSpace{}%
\AgdaSymbol{\AgdaUnderscore{}}\AgdaSpace{}%
\AgdaOperator{\AgdaInductiveConstructor{∷}}\AgdaSpace{}%
\AgdaNumber{1}\AgdaSpace{}%
\AgdaOperator{\AgdaInductiveConstructor{bounded}}\AgdaSpace{}%
\AgdaSymbol{\AgdaUnderscore{}}\AgdaSpace{}%
\AgdaOperator{\AgdaInductiveConstructor{∷}}\AgdaSpace{}%
\AgdaInductiveConstructor{[]}\AgdaSymbol{))}\AgdaSpace{}%
\AgdaSymbol{→}\AgdaSpace{}%
\AgdaNumber{42}\AgdaSpace{}%
\AgdaSymbol{;}\AgdaSpace{}%
\AgdaSymbol{\AgdaUnderscore{}}\AgdaSpace{}%
\AgdaSymbol{→}\AgdaSpace{}%
\AgdaNumber{0}\AgdaSymbol{\}}\<%
\end{code}
We defined a scalar with a value 42; a vector of 5 elements with the value
42 at the index zero and value 0 elsewhere; a matrix with the value 42 at
index $[1,1]$ and zeroes elsewhere.

For a more realistic example consider a level-polymorphic array addition.
\begin{code}%
\>[0]\AgdaFunction{plus}\AgdaSpace{}%
\AgdaSymbol{:}\AgdaSpace{}%
\AgdaSymbol{∀}\AgdaSpace{}%
\AgdaSymbol{\{}\AgdaBound{l}\AgdaSpace{}%
\AgdaBound{s}\AgdaSymbol{\}}\AgdaSpace{}%
\AgdaSymbol{→}\AgdaSpace{}%
\AgdaDatatype{Ar}\AgdaSpace{}%
\AgdaBound{l}\AgdaSpace{}%
\AgdaDatatype{ℕ}\AgdaSpace{}%
\AgdaBound{s}\AgdaSpace{}%
\AgdaSymbol{→}\AgdaSpace{}%
\AgdaDatatype{Ar}\AgdaSpace{}%
\AgdaBound{l}\AgdaSpace{}%
\AgdaDatatype{ℕ}\AgdaSpace{}%
\AgdaBound{s}\AgdaSpace{}%
\AgdaSymbol{→}\AgdaSpace{}%
\AgdaDatatype{Ar}\AgdaSpace{}%
\AgdaBound{l}\AgdaSpace{}%
\AgdaDatatype{ℕ}\AgdaSpace{}%
\AgdaBound{s}\<%
\\
\>[0]\AgdaFunction{plus}\AgdaSpace{}%
\AgdaSymbol{(}\AgdaInductiveConstructor{imap}\AgdaSpace{}%
\AgdaBound{a}\AgdaSymbol{)}\AgdaSpace{}%
\AgdaSymbol{(}\AgdaInductiveConstructor{imap}\AgdaSpace{}%
\AgdaBound{b}\AgdaSymbol{)}\AgdaSpace{}%
\AgdaSymbol{=}\AgdaSpace{}%
\AgdaInductiveConstructor{imap}\AgdaSpace{}%
\AgdaSymbol{λ}\AgdaSpace{}%
\AgdaBound{iv}\AgdaSpace{}%
\AgdaSymbol{→}\AgdaSpace{}%
\AgdaBound{a}\AgdaSpace{}%
\AgdaBound{iv}\AgdaSpace{}%
\AgdaOperator{\AgdaPrimitive{+}}\AgdaSpace{}%
\AgdaBound{b}\AgdaSpace{}%
\AgdaBound{iv}\<%
\end{code}
Generally speaking, any element-level function can be lifted to the level
of arrays: Ar with a fixed level and shape is a functor.
\begin{code}[hide]%
\>[0]\AgdaPostulate{sum}\AgdaSpace{}%
\AgdaSymbol{:}\AgdaSpace{}%
\AgdaSymbol{∀}\AgdaSpace{}%
\AgdaSymbol{\{}\AgdaBound{l}\AgdaSpace{}%
\AgdaBound{s}\AgdaSymbol{\}}\AgdaSpace{}%
\AgdaSymbol{→}\AgdaSpace{}%
\AgdaDatatype{Ar}\AgdaSpace{}%
\AgdaBound{l}\AgdaSpace{}%
\AgdaDatatype{ℕ}\AgdaSpace{}%
\AgdaBound{s}\AgdaSpace{}%
\AgdaSymbol{→}\AgdaSpace{}%
\AgdaDatatype{ℕ}\<%
\end{code}
As another realistic example consider matrix multiplication.
We formulate it in a way so that it operates on 2-dimensional arrays of
any size, given that the outer dimension of the first matrix is identical
to the inner dimension of the second one.
\begin{code}%
\>[0]\AgdaFunction{matmul}\AgdaSpace{}%
\AgdaSymbol{:}%
\>[340I]\AgdaSymbol{∀}\AgdaSpace{}%
\AgdaSymbol{\{}\AgdaBound{m}\AgdaSpace{}%
\AgdaBound{n}\AgdaSpace{}%
\AgdaBound{p}\AgdaSymbol{\}}\AgdaSpace{}%
\AgdaSymbol{→}\AgdaSpace{}%
\AgdaKeyword{let}\AgdaSpace{}%
\AgdaBound{Sh}\AgdaSpace{}%
\AgdaBound{x}\AgdaSpace{}%
\AgdaBound{y}\AgdaSpace{}%
\AgdaSymbol{=}\AgdaSpace{}%
\AgdaSymbol{((\AgdaUnderscore{}}\AgdaSpace{}%
\AgdaOperator{\AgdaInductiveConstructor{,}}\AgdaSpace{}%
\AgdaNumber{2}\AgdaSpace{}%
\AgdaOperator{\AgdaInductiveConstructor{∷}}\AgdaSpace{}%
\AgdaInductiveConstructor{[]}\AgdaSymbol{)}\AgdaSpace{}%
\AgdaOperator{\AgdaInductiveConstructor{,}}\AgdaSpace{}%
\AgdaBound{x}\AgdaSpace{}%
\AgdaOperator{\AgdaInductiveConstructor{∷}}\AgdaSpace{}%
\AgdaBound{y}\AgdaSpace{}%
\AgdaOperator{\AgdaInductiveConstructor{∷}}\AgdaSpace{}%
\AgdaInductiveConstructor{[]}\AgdaSymbol{)}\AgdaSpace{}%
\AgdaKeyword{in}\<%
\\
\>[.][@{}l@{}]\<[340I]%
\>[9]\AgdaDatatype{Ar}\AgdaSpace{}%
\AgdaNumber{2}\AgdaSpace{}%
\AgdaDatatype{ℕ}\AgdaSpace{}%
\AgdaOperator{\AgdaFunction{\$}}\AgdaSpace{}%
\AgdaBound{Sh}\AgdaSpace{}%
\AgdaBound{m}\AgdaSpace{}%
\AgdaBound{p}\AgdaSpace{}%
\AgdaSymbol{→}\AgdaSpace{}%
\AgdaDatatype{Ar}\AgdaSpace{}%
\AgdaNumber{2}\AgdaSpace{}%
\AgdaDatatype{ℕ}\AgdaSpace{}%
\AgdaOperator{\AgdaFunction{\$}}\AgdaSpace{}%
\AgdaBound{Sh}\AgdaSpace{}%
\AgdaBound{p}\AgdaSpace{}%
\AgdaBound{n}\AgdaSpace{}%
\AgdaSymbol{→}\AgdaSpace{}%
\AgdaDatatype{Ar}\AgdaSpace{}%
\AgdaNumber{2}\AgdaSpace{}%
\AgdaDatatype{ℕ}\AgdaSpace{}%
\AgdaOperator{\AgdaFunction{\$}}\AgdaSpace{}%
\AgdaBound{Sh}\AgdaSpace{}%
\AgdaBound{m}\AgdaSpace{}%
\AgdaBound{n}\<%
\\
\>[0]\AgdaFunction{matmul}\AgdaSpace{}%
\AgdaSymbol{\{}\AgdaArgument{p}\AgdaSpace{}%
\AgdaSymbol{=}\AgdaSpace{}%
\AgdaBound{p}\AgdaSymbol{\}}\AgdaSpace{}%
\AgdaSymbol{(}\AgdaInductiveConstructor{imap}\AgdaSpace{}%
\AgdaBound{a}\AgdaSymbol{)}\AgdaSpace{}%
\AgdaSymbol{(}\AgdaInductiveConstructor{imap}\AgdaSpace{}%
\AgdaBound{b}\AgdaSymbol{)}\AgdaSpace{}%
\AgdaSymbol{=}\AgdaSpace{}%
\AgdaInductiveConstructor{imap}\AgdaSpace{}%
\AgdaFunction{mat-content}\<%
\\
\>[0][@{}l@{\AgdaIndent{0}}]%
\>[2]\AgdaKeyword{where}%
\>[394I]\AgdaFunction{mat-content}\AgdaSpace{}%
\AgdaSymbol{:}\AgdaSpace{}%
\AgdaSymbol{\AgdaUnderscore{}}\<%
\\
\>[.][@{}l@{}]\<[394I]%
\>[8]\AgdaFunction{mat-content}\AgdaSpace{}%
\AgdaSymbol{(}\AgdaInductiveConstructor{ix}\AgdaSpace{}%
\AgdaSymbol{(}\AgdaBound{i}\AgdaSpace{}%
\AgdaOperator{\AgdaInductiveConstructor{∷}}\AgdaSpace{}%
\AgdaBound{j}\AgdaSpace{}%
\AgdaOperator{\AgdaInductiveConstructor{∷}}\AgdaSpace{}%
\AgdaInductiveConstructor{[]}\AgdaSymbol{))}\AgdaSpace{}%
\AgdaSymbol{=}\AgdaSpace{}%
\AgdaKeyword{let}\<%
\\
\>[8][@{}l@{\AgdaIndent{0}}]%
\>[11]\AgdaBound{t}\AgdaSpace{}%
\AgdaSymbol{:}\AgdaSpace{}%
\AgdaDatatype{Ar}\AgdaSpace{}%
\AgdaNumber{1}\AgdaSpace{}%
\AgdaDatatype{ℕ}\AgdaSpace{}%
\AgdaSymbol{(\AgdaUnderscore{}}\AgdaSpace{}%
\AgdaOperator{\AgdaInductiveConstructor{,}}\AgdaSpace{}%
\AgdaBound{p}\AgdaSpace{}%
\AgdaOperator{\AgdaInductiveConstructor{∷}}\AgdaSpace{}%
\AgdaInductiveConstructor{[]}\AgdaSymbol{)}\<%
\\
\>[11]\AgdaBound{t}\AgdaSpace{}%
\AgdaSymbol{=}\AgdaSpace{}%
\AgdaInductiveConstructor{imap}\AgdaSpace{}%
\AgdaSymbol{λ}\AgdaSpace{}%
\AgdaSymbol{\{(}\AgdaInductiveConstructor{ix}\AgdaSpace{}%
\AgdaSymbol{(}\AgdaBound{k}\AgdaSpace{}%
\AgdaOperator{\AgdaInductiveConstructor{∷}}\AgdaSpace{}%
\AgdaInductiveConstructor{[]}\AgdaSymbol{))}\AgdaSpace{}%
\AgdaSymbol{→}\AgdaSpace{}%
\AgdaBound{a}\AgdaSpace{}%
\AgdaSymbol{(}\AgdaInductiveConstructor{ix}\AgdaSpace{}%
\AgdaOperator{\AgdaFunction{\$}}\AgdaSpace{}%
\AgdaBound{i}\AgdaSpace{}%
\AgdaOperator{\AgdaInductiveConstructor{∷}}\AgdaSpace{}%
\AgdaBound{k}\AgdaSpace{}%
\AgdaOperator{\AgdaInductiveConstructor{∷}}\AgdaSpace{}%
\AgdaInductiveConstructor{[]}\AgdaSymbol{)}\AgdaSpace{}%
\AgdaOperator{\AgdaPrimitive{*}}\AgdaSpace{}%
\AgdaBound{b}\AgdaSpace{}%
\AgdaSymbol{(}\AgdaInductiveConstructor{ix}\AgdaSpace{}%
\AgdaOperator{\AgdaFunction{\$}}\AgdaSpace{}%
\AgdaBound{k}\AgdaSpace{}%
\AgdaOperator{\AgdaInductiveConstructor{∷}}\AgdaSpace{}%
\AgdaBound{j}\AgdaSpace{}%
\AgdaOperator{\AgdaInductiveConstructor{∷}}\AgdaSpace{}%
\AgdaInductiveConstructor{[]}\AgdaSymbol{)\}}\<%
\\
\>[11]\AgdaKeyword{in}\AgdaSpace{}%
\AgdaPostulate{sum}\AgdaSpace{}%
\AgdaBound{t}\<%
\end{code}


\section{Practical Applications}
\begin{code}[hide]%
\>[0]\AgdaComment{\ {-}{-}\{-\# OPTIONS --rewriting  \#-\}}\<%
\\
\>[0]\AgdaKeyword{open}\AgdaSpace{}%
\AgdaKeyword{import}\AgdaSpace{}%
\AgdaModule{Data.Nat}\<%
\\
\>[0]\AgdaKeyword{open}\AgdaSpace{}%
\AgdaKeyword{import}\AgdaSpace{}%
\AgdaModule{Data.Nat.DivMod}\<%
\\
\>[0]\AgdaKeyword{open}\AgdaSpace{}%
\AgdaKeyword{import}\AgdaSpace{}%
\AgdaModule{Data.Nat.Properties}\<%
\\
\\[\AgdaEmptyExtraSkip]%
\>[0]\AgdaKeyword{open}\AgdaSpace{}%
\AgdaKeyword{import}\AgdaSpace{}%
\AgdaModule{Data.Product}\AgdaSpace{}%
\AgdaKeyword{hiding}\AgdaSpace{}%
\AgdaSymbol{(}\AgdaFunction{map}\AgdaSymbol{)}\<%
\\
\>[0]\AgdaKeyword{open}\AgdaSpace{}%
\AgdaKeyword{import}\AgdaSpace{}%
\AgdaModule{Data.Sum}\AgdaSpace{}%
\AgdaKeyword{hiding}\AgdaSpace{}%
\AgdaSymbol{(}\AgdaFunction{map}\AgdaSymbol{)}\<%
\\
\>[0]\AgdaKeyword{open}\AgdaSpace{}%
\AgdaKeyword{import}\AgdaSpace{}%
\AgdaModule{Function}\AgdaSpace{}%
\AgdaKeyword{using}\AgdaSpace{}%
\AgdaSymbol{(}\AgdaOperator{\AgdaFunction{\AgdaUnderscore{}\$\AgdaUnderscore{}}}\AgdaSpace{}%
\AgdaSymbol{;}\AgdaSpace{}%
\AgdaOperator{\AgdaFunction{\AgdaUnderscore{}∘\AgdaUnderscore{}}}\AgdaSpace{}%
\AgdaSymbol{;}\AgdaSpace{}%
\AgdaOperator{\AgdaFunction{case\AgdaUnderscore{}of\AgdaUnderscore{}}}\AgdaSymbol{)}\<%
\\
\>[0]\AgdaKeyword{open}\AgdaSpace{}%
\AgdaKeyword{import}\AgdaSpace{}%
\AgdaModule{Data.Vec}\AgdaSpace{}%
\AgdaKeyword{hiding}\AgdaSpace{}%
\AgdaSymbol{(}\AgdaFunction{sum}\AgdaSymbol{;}\AgdaSpace{}%
\AgdaFunction{map}\AgdaSymbol{)}\<%
\\
\>[0]\AgdaKeyword{open}\AgdaSpace{}%
\AgdaKeyword{import}\AgdaSpace{}%
\AgdaModule{Data.Vec.Properties}\<%
\\
\>[0]\AgdaKeyword{open}\AgdaSpace{}%
\AgdaKeyword{import}\AgdaSpace{}%
\AgdaModule{Data.Unit}\AgdaSpace{}%
\AgdaKeyword{hiding}\AgdaSpace{}%
\AgdaSymbol{(}\AgdaOperator{\AgdaFunction{\AgdaUnderscore{}≟\AgdaUnderscore{}}}\AgdaSymbol{;}\AgdaSpace{}%
\AgdaOperator{\AgdaRecord{\AgdaUnderscore{}≤\AgdaUnderscore{}}}\AgdaSymbol{)}\<%
\\
\\[\AgdaEmptyExtraSkip]%
\>[0]\AgdaKeyword{open}\AgdaSpace{}%
\AgdaKeyword{import}\AgdaSpace{}%
\AgdaModule{Relation.Nullary}\<%
\\
\>[0]\AgdaKeyword{open}\AgdaSpace{}%
\AgdaKeyword{import}\AgdaSpace{}%
\AgdaModule{Relation.Nullary.Decidable}\AgdaSpace{}%
\AgdaKeyword{hiding}\AgdaSpace{}%
\AgdaSymbol{(}\AgdaFunction{map}\AgdaSymbol{)}\<%
\\
\>[0]\AgdaKeyword{open}\AgdaSpace{}%
\AgdaKeyword{import}\AgdaSpace{}%
\AgdaModule{Relation.Nullary.Negation}\<%
\\
\>[0]\AgdaKeyword{open}\AgdaSpace{}%
\AgdaKeyword{import}\AgdaSpace{}%
\AgdaModule{Relation.Binary.PropositionalEquality}\<%
\\
\>[0]\AgdaKeyword{open}\AgdaSpace{}%
\AgdaKeyword{import}\AgdaSpace{}%
\AgdaModule{Relation.Binary.Core}\<%
\\
\>[0]\AgdaKeyword{open}\AgdaSpace{}%
\AgdaKeyword{import}\AgdaSpace{}%
\AgdaModule{Relation.Binary}\<%
\\
\\[\AgdaEmptyExtraSkip]%
\>[0]\AgdaKeyword{open}\AgdaSpace{}%
\AgdaKeyword{import}\AgdaSpace{}%
\AgdaModule{Data.Empty}\<%
\\
\>[0]\AgdaKeyword{open}\AgdaSpace{}%
\AgdaKeyword{import}\AgdaSpace{}%
\AgdaModule{Data.Nat.Divisibility}\<%
\\
\\[\AgdaEmptyExtraSkip]%
\>[0]\AgdaKeyword{open}\AgdaSpace{}%
\AgdaKeyword{import}\AgdaSpace{}%
\AgdaModule{Data.Nat.Solver}\<%
\\
\>[0]\AgdaKeyword{open}\AgdaSpace{}%
\AgdaModule{+-*-Solver}\<%
\\
\\[\AgdaEmptyExtraSkip]%
\>[0]\AgdaKeyword{instance}\<%
\\
\>[0][@{}l@{\AgdaIndent{0}}]%
\>[2]\AgdaFunction{auto≥}\AgdaSpace{}%
\AgdaSymbol{:}\AgdaSpace{}%
\AgdaSymbol{∀}\AgdaSpace{}%
\AgdaSymbol{\{}\AgdaBound{m}\AgdaSpace{}%
\AgdaBound{n}\AgdaSpace{}%
\AgdaSymbol{:}\AgdaSpace{}%
\AgdaDatatype{ℕ}\AgdaSymbol{\}}\AgdaSpace{}%
\AgdaSymbol{→}\AgdaSpace{}%
\AgdaSymbol{\{\{}\AgdaBound{\AgdaUnderscore{}}\AgdaSpace{}%
\AgdaSymbol{:}\AgdaSpace{}%
\AgdaFunction{True}\AgdaSpace{}%
\AgdaSymbol{(}\AgdaBound{m}\AgdaSpace{}%
\AgdaOperator{\AgdaFunction{≥?}}\AgdaSpace{}%
\AgdaBound{n}\AgdaSymbol{)\}\}}\AgdaSpace{}%
\AgdaSymbol{→}\AgdaSpace{}%
\AgdaBound{m}\AgdaSpace{}%
\AgdaOperator{\AgdaFunction{≥}}\AgdaSpace{}%
\AgdaBound{n}\<%
\\
\>[2]\AgdaFunction{auto≥}\AgdaSpace{}%
\AgdaSymbol{\{}\AgdaBound{m}\AgdaSymbol{\}}\AgdaSpace{}%
\AgdaSymbol{\{}\AgdaBound{n}\AgdaSymbol{\}}\AgdaSpace{}%
\AgdaSymbol{\{\{}\AgdaBound{c}\AgdaSymbol{\}\}}\AgdaSpace{}%
\AgdaSymbol{=}\AgdaSpace{}%
\AgdaFunction{toWitness}\AgdaSpace{}%
\AgdaBound{c}\<%
\\
\\[\AgdaEmptyExtraSkip]%
\\[\AgdaEmptyExtraSkip]%
\\[\AgdaEmptyExtraSkip]%
\>[0]\AgdaKeyword{record}\AgdaSpace{}%
\AgdaRecord{BFin}\AgdaSpace{}%
\AgdaSymbol{(}\AgdaBound{u}\AgdaSpace{}%
\AgdaSymbol{:}\AgdaSpace{}%
\AgdaDatatype{ℕ}\AgdaSymbol{)}\AgdaSpace{}%
\AgdaSymbol{:}\AgdaSpace{}%
\AgdaPrimitiveType{Set}\AgdaSpace{}%
\AgdaKeyword{where}\<%
\\
\>[0][@{}l@{\AgdaIndent{0}}]%
\>[2]\AgdaKeyword{constructor}\AgdaSpace{}%
\AgdaOperator{\AgdaInductiveConstructor{\AgdaUnderscore{}bounded\AgdaUnderscore{}}}\<%
\\
\>[2]\AgdaKeyword{field}\<%
\\
\>[2][@{}l@{\AgdaIndent{0}}]%
\>[4]\AgdaField{v}\AgdaSpace{}%
\AgdaSymbol{:}\AgdaSpace{}%
\AgdaDatatype{ℕ}\<%
\\
\>[4]\AgdaSymbol{.}\AgdaField{v<u}\AgdaSpace{}%
\AgdaSymbol{:}\AgdaSpace{}%
\AgdaBound{v}\AgdaSpace{}%
\AgdaOperator{\AgdaFunction{<}}\AgdaSpace{}%
\AgdaBound{u}\<%
\\
\\[\AgdaEmptyExtraSkip]%
\>[0]\AgdaKeyword{open}\AgdaSpace{}%
\AgdaModule{BFin}\<%
\\
\\[\AgdaEmptyExtraSkip]%
\\[\AgdaEmptyExtraSkip]%
\>[0]\AgdaFunction{blookup}\AgdaSpace{}%
\AgdaSymbol{:}\AgdaSpace{}%
\AgdaSymbol{∀}\AgdaSpace{}%
\AgdaSymbol{\{}\AgdaBound{a}\AgdaSymbol{\}\{}\AgdaBound{n}\AgdaSymbol{\}\{}\AgdaBound{X}\AgdaSpace{}%
\AgdaSymbol{:}\AgdaSpace{}%
\AgdaPrimitiveType{Set}\AgdaSpace{}%
\AgdaBound{a}\AgdaSymbol{\}}\AgdaSpace{}%
\AgdaSymbol{→}\AgdaSpace{}%
\AgdaDatatype{Vec}\AgdaSpace{}%
\AgdaBound{X}\AgdaSpace{}%
\AgdaBound{n}\AgdaSpace{}%
\AgdaSymbol{→}\AgdaSpace{}%
\AgdaRecord{BFin}\AgdaSpace{}%
\AgdaBound{n}\AgdaSpace{}%
\AgdaSymbol{→}\AgdaSpace{}%
\AgdaBound{X}\<%
\\
\>[0]\AgdaFunction{blookup}\AgdaSpace{}%
\AgdaSymbol{\{}\AgdaArgument{n}\AgdaSpace{}%
\AgdaSymbol{=}\AgdaSpace{}%
\AgdaInductiveConstructor{suc}\AgdaSpace{}%
\AgdaBound{n}\AgdaSymbol{\}}\AgdaSpace{}%
\AgdaSymbol{(}\AgdaBound{x}\AgdaSpace{}%
\AgdaOperator{\AgdaInductiveConstructor{∷}}\AgdaSpace{}%
\AgdaBound{xs}\AgdaSymbol{)}\AgdaSpace{}%
\AgdaSymbol{(}\AgdaInductiveConstructor{zero}\AgdaSpace{}%
\AgdaOperator{\AgdaInductiveConstructor{bounded}}\AgdaSpace{}%
\AgdaBound{v<u}\AgdaSymbol{)}\AgdaSpace{}%
\AgdaSymbol{=}\AgdaSpace{}%
\AgdaBound{x}\<%
\\
\>[0]\AgdaFunction{blookup}\AgdaSpace{}%
\AgdaSymbol{\{}\AgdaArgument{n}\AgdaSpace{}%
\AgdaSymbol{=}\AgdaSpace{}%
\AgdaInductiveConstructor{suc}\AgdaSpace{}%
\AgdaBound{n}\AgdaSymbol{\}}\AgdaSpace{}%
\AgdaSymbol{(}\AgdaBound{x}\AgdaSpace{}%
\AgdaOperator{\AgdaInductiveConstructor{∷}}\AgdaSpace{}%
\AgdaBound{xs}\AgdaSymbol{)}\AgdaSpace{}%
\AgdaSymbol{(}\AgdaInductiveConstructor{suc}\AgdaSpace{}%
\AgdaBound{i}\AgdaSpace{}%
\AgdaOperator{\AgdaInductiveConstructor{bounded}}\AgdaSpace{}%
\AgdaBound{v<u}\AgdaSymbol{)}\AgdaSpace{}%
\AgdaSymbol{=}\AgdaSpace{}%
\AgdaFunction{blookup}\AgdaSpace{}%
\AgdaBound{xs}\AgdaSpace{}%
\AgdaSymbol{(}\AgdaBound{i}\AgdaSpace{}%
\AgdaOperator{\AgdaInductiveConstructor{bounded}}\AgdaSpace{}%
\AgdaFunction{≤-pred}\AgdaSpace{}%
\AgdaBound{v<u}\AgdaSymbol{)}\<%
\\
\\[\AgdaEmptyExtraSkip]%
\\[\AgdaEmptyExtraSkip]%
\>[0]\AgdaKeyword{data}\AgdaSpace{}%
\AgdaDatatype{FlatIx}\AgdaSpace{}%
\AgdaSymbol{:}\AgdaSpace{}%
\AgdaSymbol{(}\AgdaBound{d}\AgdaSpace{}%
\AgdaSymbol{:}\AgdaSpace{}%
\AgdaDatatype{ℕ}\AgdaSymbol{)}\AgdaSpace{}%
\AgdaSymbol{→}\AgdaSpace{}%
\AgdaSymbol{(}\AgdaBound{s}\AgdaSpace{}%
\AgdaSymbol{:}\AgdaSpace{}%
\AgdaDatatype{Vec}\AgdaSpace{}%
\AgdaDatatype{ℕ}\AgdaSpace{}%
\AgdaBound{d}\AgdaSymbol{)}\AgdaSpace{}%
\AgdaSymbol{→}\AgdaSpace{}%
\AgdaPrimitiveType{Set}\AgdaSpace{}%
\AgdaKeyword{where}\<%
\\
\>[0][@{}l@{\AgdaIndent{0}}]%
\>[2]\AgdaInductiveConstructor{[]}%
\>[7]\AgdaSymbol{:}\AgdaSpace{}%
\AgdaDatatype{FlatIx}\AgdaSpace{}%
\AgdaNumber{0}\AgdaSpace{}%
\AgdaInductiveConstructor{[]}\<%
\\
\>[2]\AgdaOperator{\AgdaInductiveConstructor{\AgdaUnderscore{}∷\AgdaUnderscore{}}}\AgdaSpace{}%
\AgdaSymbol{:}\AgdaSpace{}%
\AgdaSymbol{∀}\AgdaSpace{}%
\AgdaSymbol{\{}\AgdaBound{d}\AgdaSpace{}%
\AgdaBound{s}\AgdaSpace{}%
\AgdaBound{x}\AgdaSymbol{\}}\AgdaSpace{}%
\AgdaSymbol{→}\AgdaSpace{}%
\AgdaRecord{BFin}\AgdaSpace{}%
\AgdaBound{x}\AgdaSpace{}%
\AgdaSymbol{→}\AgdaSpace{}%
\AgdaSymbol{(}\AgdaBound{ix}\AgdaSpace{}%
\AgdaSymbol{:}\AgdaSpace{}%
\AgdaDatatype{FlatIx}\AgdaSpace{}%
\AgdaBound{d}\AgdaSpace{}%
\AgdaBound{s}\AgdaSymbol{)}\AgdaSpace{}%
\AgdaSymbol{→}\AgdaSpace{}%
\AgdaDatatype{FlatIx}\AgdaSpace{}%
\AgdaSymbol{(}\AgdaInductiveConstructor{suc}\AgdaSpace{}%
\AgdaBound{d}\AgdaSymbol{)}\AgdaSpace{}%
\AgdaSymbol{(}\AgdaBound{x}\AgdaSpace{}%
\AgdaOperator{\AgdaInductiveConstructor{∷}}\AgdaSpace{}%
\AgdaBound{s}\AgdaSymbol{)}\<%
\\
\\[\AgdaEmptyExtraSkip]%
\\[\AgdaEmptyExtraSkip]%
\>[0]\AgdaFunction{flat-prod}\AgdaSpace{}%
\AgdaSymbol{:}\AgdaSpace{}%
\AgdaSymbol{∀}\AgdaSpace{}%
\AgdaSymbol{\{}\AgdaBound{n}\AgdaSymbol{\}}\AgdaSpace{}%
\AgdaSymbol{→}\AgdaSpace{}%
\AgdaDatatype{Vec}\AgdaSpace{}%
\AgdaDatatype{ℕ}\AgdaSpace{}%
\AgdaBound{n}\AgdaSpace{}%
\AgdaSymbol{→}\AgdaSpace{}%
\AgdaDatatype{ℕ}\<%
\\
\>[0]\AgdaFunction{flat-prod}\AgdaSpace{}%
\AgdaSymbol{=}\AgdaSpace{}%
\AgdaFunction{foldr}\AgdaSpace{}%
\AgdaSymbol{\AgdaUnderscore{}}\AgdaSpace{}%
\AgdaOperator{\AgdaPrimitive{\AgdaUnderscore{}*\AgdaUnderscore{}}}\AgdaSpace{}%
\AgdaNumber{1}\<%
\\
\\[\AgdaEmptyExtraSkip]%
\>[0]\AgdaComment{\ {-}{-} The type of shape of arrays of level `l`}\<%
\\
\>[0]\AgdaFunction{ShType}\AgdaSpace{}%
\AgdaSymbol{:}\AgdaSpace{}%
\AgdaSymbol{(}\AgdaBound{l}\AgdaSpace{}%
\AgdaSymbol{:}\AgdaSpace{}%
\AgdaDatatype{ℕ}\AgdaSymbol{)}\AgdaSpace{}%
\AgdaSymbol{→}\AgdaSpace{}%
\AgdaPrimitiveType{Set}\<%
\\
\\[\AgdaEmptyExtraSkip]%
\>[0]\AgdaComment{\ {-}{-} The type of indices into arrays of level `l`}\<%
\\
\>[0]\AgdaFunction{IxType}\AgdaSpace{}%
\AgdaSymbol{:}\AgdaSpace{}%
\AgdaSymbol{(}\AgdaBound{l}\AgdaSpace{}%
\AgdaSymbol{:}\AgdaSpace{}%
\AgdaDatatype{ℕ}\AgdaSymbol{)}\AgdaSpace{}%
\AgdaSymbol{→}\AgdaSpace{}%
\AgdaFunction{ShType}\AgdaSpace{}%
\AgdaBound{l}\AgdaSpace{}%
\AgdaSymbol{→}\AgdaSpace{}%
\AgdaPrimitiveType{Set}\<%
\\
\\[\AgdaEmptyExtraSkip]%
\>[0]\AgdaComment{\ {-}{-} Representaiton for an array of level `l`.}\<%
\\
\>[0]\AgdaComment{\ {-}{-} FIXME: we have to explain what are the properties of}\<%
\\
\>[0]\AgdaComment{\ {-}{-} the representation, and when the representation is adequate.}\<%
\\
\>[0]\AgdaFunction{ReprAr}\AgdaSpace{}%
\AgdaSymbol{:}\AgdaSpace{}%
\AgdaSymbol{∀}\AgdaSpace{}%
\AgdaBound{l}\AgdaSpace{}%
\AgdaSymbol{(}\AgdaBound{X}\AgdaSpace{}%
\AgdaSymbol{:}\AgdaSpace{}%
\AgdaPrimitiveType{Set}\AgdaSymbol{)}\AgdaSpace{}%
\AgdaSymbol{→}\AgdaSpace{}%
\AgdaPrimitiveType{Set}\<%
\\
\\[\AgdaEmptyExtraSkip]%
\>[0]\AgdaComment{\ {-}{-} Representation of indices into arrays of level `l`}\<%
\\
\>[0]\AgdaKeyword{record}\AgdaSpace{}%
\AgdaRecord{Ix}\AgdaSpace{}%
\AgdaSymbol{(}\AgdaBound{l}\AgdaSpace{}%
\AgdaSymbol{:}\AgdaSpace{}%
\AgdaDatatype{ℕ}\AgdaSymbol{)}\AgdaSpace{}%
\AgdaSymbol{(}\AgdaBound{s}\AgdaSpace{}%
\AgdaSymbol{:}\AgdaSpace{}%
\AgdaFunction{ShType}\AgdaSpace{}%
\AgdaBound{l}\AgdaSymbol{)}\AgdaSpace{}%
\AgdaSymbol{:}\AgdaSpace{}%
\AgdaPrimitiveType{Set}\AgdaSpace{}%
\AgdaKeyword{where}\<%
\\
\>[0][@{}l@{\AgdaIndent{0}}]%
\>[2]\AgdaComment{\ {-}{-} In this particular case, additionally to the index}\<%
\\
\>[2]\AgdaComment{\ {-}{-} representation we carry the type argument carries the}\<%
\\
\>[2]\AgdaComment{\ {-}{-} shape of the index space where this index is coming from.}\<%
\\
\>[2]\AgdaComment{\ {-}{-} We need this to distinguish the indices from diffrent index}\<%
\\
\>[2]\AgdaComment{\ {-}{-} spaces that have the same representation.}\<%
\\
\>[2]\AgdaKeyword{constructor}\AgdaSpace{}%
\AgdaInductiveConstructor{ix}\<%
\\
\>[2]\AgdaKeyword{field}\<%
\\
\>[2][@{}l@{\AgdaIndent{0}}]%
\>[4]\AgdaField{flat-ix}\AgdaSpace{}%
\AgdaSymbol{:}\AgdaSpace{}%
\AgdaFunction{IxType}\AgdaSpace{}%
\AgdaBound{l}\AgdaSpace{}%
\AgdaBound{s}\<%
\\
\\[\AgdaEmptyExtraSkip]%
\>[0]\AgdaComment{\ {-}{-}}\<%
\\
\>[0]\AgdaKeyword{data}\AgdaSpace{}%
\AgdaDatatype{Ar}\AgdaSpace{}%
\AgdaSymbol{(}\AgdaBound{l}\AgdaSpace{}%
\AgdaSymbol{:}\AgdaSpace{}%
\AgdaDatatype{ℕ}\AgdaSymbol{)}\AgdaSpace{}%
\AgdaSymbol{(}\AgdaBound{X}\AgdaSpace{}%
\AgdaSymbol{:}\AgdaSpace{}%
\AgdaPrimitiveType{Set}\AgdaSymbol{)}\AgdaSpace{}%
\AgdaSymbol{(}\AgdaBound{s}\AgdaSpace{}%
\AgdaSymbol{:}\AgdaSpace{}%
\AgdaFunction{ShType}\AgdaSpace{}%
\AgdaBound{l}\AgdaSymbol{)}\AgdaSpace{}%
\AgdaSymbol{:}\AgdaSpace{}%
\AgdaPrimitiveType{Set}\AgdaSpace{}%
\AgdaKeyword{where}\<%
\\
\>[0][@{}l@{\AgdaIndent{0}}]%
\>[2]\AgdaInductiveConstructor{imap}\AgdaSpace{}%
\AgdaSymbol{:}\AgdaSpace{}%
\AgdaSymbol{(}\AgdaRecord{Ix}\AgdaSpace{}%
\AgdaBound{l}\AgdaSpace{}%
\AgdaBound{s}\AgdaSpace{}%
\AgdaSymbol{→}\AgdaSpace{}%
\AgdaBound{X}\AgdaSymbol{)}\AgdaSpace{}%
\AgdaSymbol{→}\AgdaSpace{}%
\AgdaDatatype{Ar}\AgdaSpace{}%
\AgdaBound{l}\AgdaSpace{}%
\AgdaBound{X}\AgdaSpace{}%
\AgdaBound{s}\<%
\\
\\[\AgdaEmptyExtraSkip]%
\\[\AgdaEmptyExtraSkip]%
\>[0]\AgdaFunction{prod}\AgdaSpace{}%
\AgdaSymbol{:}\AgdaSpace{}%
\AgdaSymbol{∀}\AgdaSpace{}%
\AgdaSymbol{\{}\AgdaBound{l}\AgdaSymbol{\}}\AgdaSpace{}%
\AgdaSymbol{→}\AgdaSpace{}%
\AgdaFunction{ShType}\AgdaSpace{}%
\AgdaBound{l}\AgdaSpace{}%
\AgdaSymbol{→}\AgdaSpace{}%
\AgdaDatatype{ℕ}\<%
\\
\\[\AgdaEmptyExtraSkip]%
\>[0]\AgdaComment{\ {-}{-} Note here that representation types of Shape and the}\<%
\\
\>[0]\AgdaComment{\ {-}{-} Array itself don't have to match.  It is convenient}\<%
\\
\>[0]\AgdaComment{\ {-}{-} when they do, but for more complex array types this}\<%
\\
\>[0]\AgdaComment{\ {-}{-} is not the case.}\<%
\\
\>[0]\AgdaFunction{ShType}\AgdaSpace{}%
\AgdaInductiveConstructor{zero}%
\>[15]\AgdaSymbol{=}\AgdaSpace{}%
\AgdaRecord{⊤}\<%
\\
\>[0]\AgdaFunction{ShType}\AgdaSpace{}%
\AgdaSymbol{(}\AgdaInductiveConstructor{suc}\AgdaSpace{}%
\AgdaBound{l}\AgdaSymbol{)}\AgdaSpace{}%
\AgdaSymbol{=}%
\>[267I]\AgdaFunction{ReprAr}\AgdaSpace{}%
\AgdaBound{l}\AgdaSpace{}%
\AgdaDatatype{ℕ}\<%
\\
\>[.][@{}l@{}]\<[267I]%
\>[17]\AgdaComment{\ {-}{-}Σ (ReprSh l) λ s → Vec ℕ (prod s)}\<%
\\
\\[\AgdaEmptyExtraSkip]%
\>[0]\AgdaFunction{ReprAr}\AgdaSpace{}%
\AgdaBound{l}\AgdaSpace{}%
\AgdaBound{X}\AgdaSpace{}%
\AgdaSymbol{=}\AgdaSpace{}%
\AgdaRecord{Σ}\AgdaSpace{}%
\AgdaSymbol{(}\AgdaFunction{ShType}\AgdaSpace{}%
\AgdaBound{l}\AgdaSymbol{)}\AgdaSpace{}%
\AgdaSymbol{λ}\AgdaSpace{}%
\AgdaBound{s}\AgdaSpace{}%
\AgdaSymbol{→}\AgdaSpace{}%
\AgdaDatatype{Vec}\AgdaSpace{}%
\AgdaBound{X}\AgdaSpace{}%
\AgdaSymbol{(}\AgdaFunction{prod}\AgdaSpace{}%
\AgdaSymbol{\{}\AgdaArgument{l}\AgdaSpace{}%
\AgdaSymbol{=}\AgdaSpace{}%
\AgdaBound{l}\AgdaSymbol{\}}\AgdaSpace{}%
\AgdaBound{s}\AgdaSymbol{)}\<%
\\
\\[\AgdaEmptyExtraSkip]%
\>[0]\AgdaComment{\ {-}{-} In principle, we can say that this type on its own is}\<%
\\
\>[0]\AgdaComment{\ {-}{-} a type of array indices.  In this case however, we won't}\<%
\\
\>[0]\AgdaComment{\ {-}{-} be able to distinguish between indices into different}\<%
\\
\>[0]\AgdaComment{\ {-}{-} shapes.  E.g. index ⟨2⟩ can index arrays Ar 1 X 5}\<%
\\
\>[0]\AgdaComment{\ {-}{-} Ar 2 X [5], etc.  This can be both a bug and a feature.}\<%
\\
\>[0]\AgdaFunction{IxType}\AgdaSpace{}%
\AgdaInductiveConstructor{zero}\AgdaSpace{}%
\AgdaInductiveConstructor{tt}\AgdaSpace{}%
\AgdaSymbol{=}\AgdaSpace{}%
\AgdaRecord{⊤}\<%
\\
\>[0]\AgdaFunction{IxType}\AgdaSpace{}%
\AgdaSymbol{(}\AgdaInductiveConstructor{suc}\AgdaSpace{}%
\AgdaBound{l}\AgdaSymbol{)}\AgdaSpace{}%
\AgdaSymbol{(}\AgdaBound{s}\AgdaSpace{}%
\AgdaOperator{\AgdaInductiveConstructor{,}}\AgdaSpace{}%
\AgdaBound{vec}\AgdaSymbol{)}\AgdaSpace{}%
\AgdaSymbol{=}\AgdaSpace{}%
\AgdaDatatype{FlatIx}\AgdaSpace{}%
\AgdaSymbol{(}\AgdaFunction{prod}\AgdaSpace{}%
\AgdaBound{s}\AgdaSymbol{)}\AgdaSpace{}%
\AgdaBound{vec}\<%
\\
\\[\AgdaEmptyExtraSkip]%
\\[\AgdaEmptyExtraSkip]%
\>[0]\AgdaFunction{prod}\AgdaSpace{}%
\AgdaSymbol{\{}\AgdaInductiveConstructor{zero}\AgdaSymbol{\}}%
\>[13]\AgdaBound{sh}\AgdaSpace{}%
\AgdaSymbol{=}\AgdaSpace{}%
\AgdaNumber{1}\<%
\\
\>[0]\AgdaFunction{prod}\AgdaSpace{}%
\AgdaSymbol{\{}\AgdaInductiveConstructor{suc}\AgdaSpace{}%
\AgdaBound{l}\AgdaSymbol{\}}\AgdaSpace{}%
\AgdaSymbol{(}\AgdaBound{s}\AgdaSpace{}%
\AgdaOperator{\AgdaInductiveConstructor{,}}\AgdaSpace{}%
\AgdaBound{vec}\AgdaSymbol{)}\AgdaSpace{}%
\AgdaSymbol{=}\AgdaSpace{}%
\AgdaFunction{flat-prod}\AgdaSpace{}%
\AgdaBound{vec}\<%
\\
\\[\AgdaEmptyExtraSkip]%
\>[0]\AgdaFunction{unimap}\AgdaSpace{}%
\AgdaSymbol{:}\AgdaSpace{}%
\AgdaSymbol{∀}\AgdaSpace{}%
\AgdaSymbol{\{}\AgdaBound{l}\AgdaSpace{}%
\AgdaBound{X}\AgdaSpace{}%
\AgdaBound{s}\AgdaSymbol{\}}\AgdaSpace{}%
\AgdaSymbol{→}\AgdaSpace{}%
\AgdaDatatype{Ar}\AgdaSpace{}%
\AgdaBound{l}\AgdaSpace{}%
\AgdaBound{X}\AgdaSpace{}%
\AgdaBound{s}\AgdaSpace{}%
\AgdaSymbol{→}\AgdaSpace{}%
\AgdaSymbol{\AgdaUnderscore{}}\<%
\\
\>[0]\AgdaFunction{unimap}\AgdaSpace{}%
\AgdaSymbol{(}\AgdaInductiveConstructor{imap}\AgdaSpace{}%
\AgdaBound{a}\AgdaSymbol{)}\AgdaSpace{}%
\AgdaSymbol{=}\AgdaSpace{}%
\AgdaBound{a}\<%
\\
\\[\AgdaEmptyExtraSkip]%
\>[0]\AgdaPostulate{repr}\AgdaSpace{}%
\AgdaSymbol{:}\AgdaSpace{}%
\AgdaSymbol{∀}\AgdaSpace{}%
\AgdaSymbol{\{}\AgdaBound{X}\AgdaSpace{}%
\AgdaBound{l}\AgdaSpace{}%
\AgdaBound{s}\AgdaSymbol{\}}\AgdaSpace{}%
\AgdaSymbol{→}\AgdaSpace{}%
\AgdaDatatype{Ar}\AgdaSpace{}%
\AgdaBound{l}\AgdaSpace{}%
\AgdaBound{X}\AgdaSpace{}%
\AgdaBound{s}\AgdaSpace{}%
\AgdaSymbol{→}\AgdaSpace{}%
\AgdaFunction{ReprAr}\AgdaSpace{}%
\AgdaBound{l}\AgdaSpace{}%
\AgdaBound{X}\<%
\\
\\[\AgdaEmptyExtraSkip]%
\>[0]\AgdaFunction{sum}\AgdaSpace{}%
\AgdaSymbol{:}\AgdaSpace{}%
\AgdaSymbol{∀}\AgdaSpace{}%
\AgdaSymbol{\{}\AgdaBound{l}\AgdaSpace{}%
\AgdaBound{s}\AgdaSymbol{\}}\AgdaSpace{}%
\AgdaSymbol{→}\AgdaSpace{}%
\AgdaDatatype{Ar}\AgdaSpace{}%
\AgdaBound{l}\AgdaSpace{}%
\AgdaDatatype{ℕ}\AgdaSpace{}%
\AgdaBound{s}\AgdaSpace{}%
\AgdaSymbol{→}\AgdaSpace{}%
\AgdaDatatype{ℕ}\<%
\\
\>[0]\AgdaFunction{sum}\AgdaSpace{}%
\AgdaSymbol{\{}\AgdaBound{l}\AgdaSymbol{\}}\AgdaSpace{}%
\AgdaBound{a}\AgdaSpace{}%
\AgdaSymbol{=}\AgdaSpace{}%
\AgdaKeyword{let}\AgdaSpace{}%
\AgdaBound{s}\AgdaSpace{}%
\AgdaOperator{\AgdaInductiveConstructor{,}}\AgdaSpace{}%
\AgdaBound{v}\AgdaSpace{}%
\AgdaSymbol{=}\AgdaSpace{}%
\AgdaPostulate{repr}\AgdaSpace{}%
\AgdaBound{a}\AgdaSpace{}%
\AgdaKeyword{in}\AgdaSpace{}%
\AgdaFunction{Data.Vec.sum}\AgdaSpace{}%
\AgdaBound{v}\<%
\end{code}

Most of the typical array-based problems use arrays up to level two.
Therefore, it is reasonable to ask whether there are any applications
where availability of higher levels is handy.  As a motivating example,
we consider an instance of the average pooling problem~\cite{pooling} that is commonly used
in machine learning applications.  Given a matrix of size $(2m)×(2n)$ we
produce an $m×n$ matrix by averaging $2×2$ subarrays, for example:
\[
avgp \left(
\begin{matrix}
1 & 2 & 5 & 6 \\
3 & 4 & 7 & 8
\end{matrix}
\right)
=
\left(
\begin{matrix}
(1+2+3+4)/4 & (5+6+7+8)/4
\end{matrix}
\right)
\]

A $2×4$ matrix is turned into a $1×2$ one where all the $2×2$ subarrays
are averaged.

Even though it is straight-forward to implement this function directly:
\begin{code}[hide]%
\>[0]\AgdaKeyword{postulate}\<%
\\
\>[0][@{}l@{\AgdaIndent{0}}]%
\>[2]\AgdaPostulate{i+i'<m*2}\AgdaSpace{}%
\AgdaSymbol{:}\AgdaSpace{}%
\AgdaSymbol{∀}\AgdaSpace{}%
\AgdaSymbol{\{}\AgdaBound{X}\AgdaSpace{}%
\AgdaSymbol{:}\AgdaSpace{}%
\AgdaPrimitiveType{Set}\AgdaSymbol{\}}\AgdaSpace{}%
\AgdaSymbol{→}\AgdaSpace{}%
\AgdaBound{X}\<%
\\
\>[2]\AgdaPostulate{j+j'<n*2}\AgdaSpace{}%
\AgdaSymbol{:}\AgdaSpace{}%
\AgdaSymbol{∀}\AgdaSpace{}%
\AgdaSymbol{\{}\AgdaBound{X}\AgdaSpace{}%
\AgdaSymbol{:}\AgdaSpace{}%
\AgdaPrimitiveType{Set}\AgdaSymbol{\}}\AgdaSpace{}%
\AgdaSymbol{→}\AgdaSpace{}%
\AgdaBound{X}\<%
\end{code}
\begin{code}%
\>[0]\AgdaFunction{avgp-direct}\AgdaSpace{}%
\AgdaSymbol{:}\AgdaSpace{}%
\AgdaSymbol{∀}\AgdaSpace{}%
\AgdaSymbol{\{}\AgdaBound{m}\AgdaSpace{}%
\AgdaBound{n}\AgdaSymbol{\}}%
\>[383I]\AgdaSymbol{→}\AgdaSpace{}%
\AgdaDatatype{Ar}\AgdaSpace{}%
\AgdaNumber{2}\AgdaSpace{}%
\AgdaDatatype{ℕ}\AgdaSpace{}%
\AgdaSymbol{((}\AgdaInductiveConstructor{tt}\AgdaSpace{}%
\AgdaOperator{\AgdaInductiveConstructor{,}}\AgdaSpace{}%
\AgdaNumber{2}\AgdaSpace{}%
\AgdaOperator{\AgdaInductiveConstructor{∷}}\AgdaSpace{}%
\AgdaInductiveConstructor{[]}\AgdaSymbol{)}\AgdaSpace{}%
\AgdaOperator{\AgdaInductiveConstructor{,}}\AgdaSpace{}%
\AgdaBound{m}\AgdaSpace{}%
\AgdaOperator{\AgdaPrimitive{*}}\AgdaSpace{}%
\AgdaNumber{2}\AgdaSpace{}%
\AgdaOperator{\AgdaInductiveConstructor{∷}}\AgdaSpace{}%
\AgdaBound{n}\AgdaSpace{}%
\AgdaOperator{\AgdaPrimitive{*}}\AgdaSpace{}%
\AgdaNumber{2}\AgdaSpace{}%
\AgdaOperator{\AgdaInductiveConstructor{∷}}\AgdaSpace{}%
\AgdaInductiveConstructor{[]}\AgdaSymbol{)}\<%
\\
\>[.][@{}l@{}]\<[383I]%
\>[22]\AgdaSymbol{→}\AgdaSpace{}%
\AgdaDatatype{Ar}\AgdaSpace{}%
\AgdaNumber{2}\AgdaSpace{}%
\AgdaDatatype{ℕ}\AgdaSpace{}%
\AgdaSymbol{((}\AgdaInductiveConstructor{tt}\AgdaSpace{}%
\AgdaOperator{\AgdaInductiveConstructor{,}}\AgdaSpace{}%
\AgdaNumber{2}\AgdaSpace{}%
\AgdaOperator{\AgdaInductiveConstructor{∷}}\AgdaSpace{}%
\AgdaInductiveConstructor{[]}\AgdaSymbol{)}\AgdaSpace{}%
\AgdaOperator{\AgdaInductiveConstructor{,}}\AgdaSpace{}%
\AgdaBound{m}\AgdaSpace{}%
\AgdaOperator{\AgdaInductiveConstructor{∷}}\AgdaSpace{}%
\AgdaBound{n}\AgdaSpace{}%
\AgdaOperator{\AgdaInductiveConstructor{∷}}\AgdaSpace{}%
\AgdaInductiveConstructor{[]}\AgdaSymbol{)}\<%
\\
\>[0]\AgdaFunction{avgp-direct}\AgdaSpace{}%
\AgdaSymbol{\{}\AgdaBound{m}\AgdaSymbol{\}\{}\AgdaBound{n}\AgdaSymbol{\}}\AgdaSpace{}%
\AgdaSymbol{(}\AgdaInductiveConstructor{imap}\AgdaSpace{}%
\AgdaBound{a}\AgdaSymbol{)}\AgdaSpace{}%
\AgdaSymbol{=}\AgdaSpace{}%
\AgdaInductiveConstructor{imap}\AgdaSpace{}%
\AgdaFunction{array-content}\<%
\\
\>[0][@{}l@{\AgdaIndent{0}}]%
\>[3]\AgdaKeyword{where}%
\>[422I]\AgdaFunction{array-content}\AgdaSpace{}%
\AgdaSymbol{:}\AgdaSpace{}%
\AgdaSymbol{\AgdaUnderscore{}}\<%
\\
\>[.][@{}l@{}]\<[422I]%
\>[9]\AgdaFunction{array-content}\AgdaSpace{}%
\AgdaSymbol{(}\AgdaInductiveConstructor{ix}\AgdaSpace{}%
\AgdaSymbol{(}\AgdaBound{i}\AgdaSpace{}%
\AgdaOperator{\AgdaInductiveConstructor{bounded}}\AgdaSpace{}%
\AgdaSymbol{\AgdaUnderscore{}}\AgdaSpace{}%
\AgdaOperator{\AgdaInductiveConstructor{∷}}\AgdaSpace{}%
\AgdaBound{j}\AgdaSpace{}%
\AgdaOperator{\AgdaInductiveConstructor{bounded}}\AgdaSpace{}%
\AgdaSymbol{\AgdaUnderscore{}}\AgdaSpace{}%
\AgdaOperator{\AgdaInductiveConstructor{∷}}\AgdaSpace{}%
\AgdaInductiveConstructor{[]}\AgdaSymbol{))}\AgdaSpace{}%
\AgdaSymbol{=}\AgdaSpace{}%
\AgdaKeyword{let}\<%
\\
\>[9][@{}l@{\AgdaIndent{0}}]%
\>[11]\AgdaBound{t}\AgdaSpace{}%
\AgdaSymbol{:}\AgdaSpace{}%
\AgdaDatatype{Ar}\AgdaSpace{}%
\AgdaNumber{2}\AgdaSpace{}%
\AgdaDatatype{ℕ}\AgdaSpace{}%
\AgdaSymbol{((}\AgdaInductiveConstructor{tt}\AgdaSpace{}%
\AgdaOperator{\AgdaInductiveConstructor{,}}\AgdaSpace{}%
\AgdaNumber{2}\AgdaSpace{}%
\AgdaOperator{\AgdaInductiveConstructor{∷}}\AgdaSpace{}%
\AgdaInductiveConstructor{[]}\AgdaSymbol{)}\AgdaSpace{}%
\AgdaOperator{\AgdaInductiveConstructor{,}}\AgdaSpace{}%
\AgdaNumber{2}\AgdaSpace{}%
\AgdaOperator{\AgdaInductiveConstructor{∷}}\AgdaSpace{}%
\AgdaNumber{2}\AgdaSpace{}%
\AgdaOperator{\AgdaInductiveConstructor{∷}}\AgdaSpace{}%
\AgdaInductiveConstructor{[]}\AgdaSymbol{)}\<%
\\
\>[11]\AgdaBound{t}\AgdaSpace{}%
\AgdaSymbol{=}\AgdaSpace{}%
\AgdaInductiveConstructor{imap}\AgdaSpace{}%
\AgdaSymbol{λ}\AgdaSpace{}%
\AgdaSymbol{\{}%
\>[456I]\AgdaSymbol{(}\AgdaInductiveConstructor{ix}\AgdaSpace{}%
\AgdaSymbol{(}\AgdaBound{i'}\AgdaSpace{}%
\AgdaOperator{\AgdaInductiveConstructor{bounded}}\AgdaSpace{}%
\AgdaSymbol{\AgdaUnderscore{}}\AgdaSpace{}%
\AgdaOperator{\AgdaInductiveConstructor{∷}}\AgdaSpace{}%
\AgdaBound{j'}\AgdaSpace{}%
\AgdaOperator{\AgdaInductiveConstructor{bounded}}\AgdaSpace{}%
\AgdaSymbol{\AgdaUnderscore{}}\AgdaSpace{}%
\AgdaOperator{\AgdaInductiveConstructor{∷}}\AgdaSpace{}%
\AgdaInductiveConstructor{[]}\AgdaSymbol{))}\AgdaSpace{}%
\AgdaSymbol{→}\<%
\\
\>[.][@{}l@{}]\<[456I]%
\>[24]\AgdaBound{a}\AgdaSpace{}%
\AgdaSymbol{(}\AgdaInductiveConstructor{ix}\AgdaSpace{}%
\AgdaOperator{\AgdaFunction{\$}}\AgdaSpace{}%
\AgdaSymbol{(}\AgdaBound{i}\AgdaSpace{}%
\AgdaOperator{\AgdaPrimitive{+}}\AgdaSpace{}%
\AgdaBound{i'}\AgdaSymbol{)}\AgdaSpace{}%
\AgdaOperator{\AgdaInductiveConstructor{bounded}}\AgdaSpace{}%
\AgdaPostulate{i+i'<m*2}\AgdaSpace{}%
\AgdaOperator{\AgdaInductiveConstructor{∷}}\AgdaSpace{}%
\AgdaSymbol{(}\AgdaBound{j}\AgdaSpace{}%
\AgdaOperator{\AgdaPrimitive{+}}\AgdaSpace{}%
\AgdaBound{j'}\AgdaSymbol{)}\AgdaSpace{}%
\AgdaOperator{\AgdaInductiveConstructor{bounded}}\AgdaSpace{}%
\AgdaPostulate{j+j'<n*2}\AgdaSpace{}%
\AgdaOperator{\AgdaInductiveConstructor{∷}}\AgdaSpace{}%
\AgdaInductiveConstructor{[]}\AgdaSymbol{)}\AgdaSpace{}%
\AgdaSymbol{\}}\<%
\\
\>[11]\AgdaKeyword{in}\AgdaSpace{}%
\AgdaFunction{sum}\AgdaSpace{}%
\AgdaBound{t}\AgdaSpace{}%
\AgdaOperator{\AgdaFunction{/}}\AgdaSpace{}%
\AgdaNumber{4}\<%
\end{code}
we had to manually perform index manipulations and needed to prove
two theorems.  What if we try to implement the same algorithm using
aggregate array operations in the style of APL to avoid index manipulations?

If we were to transform the array of shape $(2m) \times (2n)$ into an array
of shape $m\times n\times 2\times 2$, we could use the concept of a
ranked operator~\cite{ranked} to apply average operation on the last two
axes.  A ranked operator can be thought of as a facility to turn a
level-2 array (\texttt{Ar 2 X $m\times n\times 2\times 2$}) into a
nested array \texttt{Ar 2 (Ar 2 X $2\times 2$) $m \times n$}.
The argument to the ranked operator typically
indicates where we ``cut'' the shape vector.

We can define a reshape operation that allows us to change the shape
of the array, given that the new shape has the same number of elements
and that the order of elements under some chosen linearisation is
preserved.  Here is the type for the reshape operation for arrays with
levels:
\begin{code}%
\>[0]\AgdaPostulate{reshape}\AgdaSpace{}%
\AgdaSymbol{:}\AgdaSpace{}%
\AgdaSymbol{∀}\AgdaSpace{}%
\AgdaSymbol{\{}\AgdaBound{X}\AgdaSpace{}%
\AgdaBound{l}\AgdaSpace{}%
\AgdaBound{l₁}\AgdaSpace{}%
\AgdaBound{s}\AgdaSpace{}%
\AgdaBound{s₁}\AgdaSymbol{\}}\AgdaSpace{}%
\AgdaSymbol{→}\AgdaSpace{}%
\AgdaDatatype{Ar}\AgdaSpace{}%
\AgdaBound{l}\AgdaSpace{}%
\AgdaBound{X}\AgdaSpace{}%
\AgdaBound{s}\AgdaSpace{}%
\AgdaSymbol{→}\AgdaSpace{}%
\AgdaFunction{prod}\AgdaSpace{}%
\AgdaBound{s}\AgdaSpace{}%
\AgdaOperator{\AgdaDatatype{≡}}\AgdaSpace{}%
\AgdaFunction{prod}\AgdaSpace{}%
\AgdaBound{s₁}\AgdaSpace{}%
\AgdaSymbol{→}\AgdaSpace{}%
\AgdaDatatype{Ar}\AgdaSpace{}%
\AgdaBound{l₁}\AgdaSpace{}%
\AgdaBound{X}\AgdaSpace{}%
\AgdaBound{s₁}\<%
\end{code}
Implementation of this operation can be found in~\cite{github}.
We choose a row-major order as our linearisation.  The reshape operation first
computes an offset into the linearised array and then turns this
offset into the index of the new shape.

Note that reshaping an array of shape $(2m)\times  (2n)$ into shape
$m\times n\times 2\times 2$ would deliver an incorrect tiling.
\[
reshape \left(
\begin{matrix}
1 & 2 & 5 & 6 \\
3 & 4 & 7 & 8
\end{matrix}
\right)
=
\left(
\begin{matrix}
      {\left(
      \begin{matrix}
        1 & 2 \\
        5 & 6
      \end{matrix}
      \right)}
      &
      {\left(
      \begin{matrix}
        3 & 4 \\
        7 & 8
      \end{matrix}
      \right)}
\end{matrix}
\right)
\quad\text{and not}\quad
\left(
\begin{matrix}
      {\left(
      \begin{matrix}
        1 & 2 \\
        3 & 4
      \end{matrix}
      \right)}
      &
      {\left(
      \begin{matrix}
        5 & 6 \\
        7 & 8
      \end{matrix}
      \right)}
\end{matrix}
\right)
\]
This is happening because under row-major order, the elements of the
row are kept together.  In order to fix this problem we need to reshape
our array into shape $m\times 2\times n\times 2$.  However, if we do so, we cannot apply
ranked operator anymore --- there is no way to ``cut'' the $m\times 2\times n\times 2$
shape vector so that $2\times 2$ become ``neighbours''.  It is here the concept
of higher-level shapes becomes useful.  If we consider the shape $m\times 2\times n\times 2$
as a linearised $2×2$ array
\(
      \left(
      \begin{matrix}
        m & 2 \\
        n & 2
      \end{matrix}
      \right)
\)
then $2\times 2$ are neighbours in the second column of the shape.  Now we
can define a smarter version of the ranked operator, that 
``cuts''
the shape across the column, and produce a level-3 nested array where
the shape of the inner array is
\(
\left(\begin{matrix}
2\\
2
\end{matrix}
\right)
\).

We make this idea precise by defining a ranked operator that ``cuts''
shapes of any levels into two parts.  Later these parts can be
used to create an element-preserving array nesting.

Let us first define the type that would capture all the allowed
``cuts'' of the given shape:
\begin{code}%
\>[0]\AgdaFunction{RankedT}\AgdaSpace{}%
\AgdaSymbol{:}\AgdaSpace{}%
\AgdaSymbol{∀}\AgdaSpace{}%
\AgdaSymbol{\{}\AgdaBound{l}\AgdaSpace{}%
\AgdaSymbol{:}\AgdaSpace{}%
\AgdaDatatype{ℕ}\AgdaSymbol{\}}\AgdaSpace{}%
\AgdaSymbol{→}\AgdaSpace{}%
\AgdaFunction{ShType}\AgdaSpace{}%
\AgdaBound{l}\AgdaSpace{}%
\AgdaSymbol{→}\AgdaSpace{}%
\AgdaPrimitiveType{Set}\<%
\\
\>[0]\AgdaFunction{RankedT}\AgdaSpace{}%
\AgdaSymbol{\{}\AgdaNumber{0}\AgdaSymbol{\}}\AgdaSpace{}%
\AgdaSymbol{\AgdaUnderscore{}}\AgdaSpace{}%
\AgdaSymbol{=}\AgdaSpace{}%
\AgdaRecord{⊤}\<%
\\
\>[0]\AgdaFunction{RankedT}\AgdaSpace{}%
\AgdaSymbol{\{}\AgdaNumber{1}\AgdaSymbol{\}}\AgdaSpace{}%
\AgdaSymbol{(}\AgdaBound{s}\AgdaSpace{}%
\AgdaOperator{\AgdaInductiveConstructor{,}}\AgdaSpace{}%
\AgdaBound{v}\AgdaSymbol{)}\AgdaSpace{}%
\AgdaSymbol{=}\AgdaSpace{}%
\AgdaRecord{BFin}\AgdaSpace{}%
\AgdaSymbol{(}\AgdaNumber{1}\AgdaSpace{}%
\AgdaOperator{\AgdaPrimitive{+}}\AgdaSpace{}%
\AgdaFunction{prod}\AgdaSpace{}%
\AgdaBound{s}\AgdaSymbol{)}\<%
\\
\>[0]\AgdaFunction{RankedT}\AgdaSpace{}%
\AgdaSymbol{\{}\AgdaInductiveConstructor{suc}\AgdaSpace{}%
\AgdaSymbol{(}\AgdaInductiveConstructor{suc}\AgdaSpace{}%
\AgdaBound{l}\AgdaSymbol{)\}}\AgdaSpace{}%
\AgdaSymbol{((}\AgdaBound{s}\AgdaSpace{}%
\AgdaOperator{\AgdaInductiveConstructor{,}}\AgdaSpace{}%
\AgdaBound{v}\AgdaSymbol{)}\AgdaSpace{}%
\AgdaOperator{\AgdaInductiveConstructor{,}}\AgdaSpace{}%
\AgdaSymbol{\AgdaUnderscore{})}\AgdaSpace{}%
\AgdaSymbol{=}\AgdaSpace{}%
\AgdaRecord{Σ}\AgdaSpace{}%
\AgdaSymbol{(}\AgdaRecord{BFin}\AgdaSpace{}%
\AgdaSymbol{(}\AgdaFunction{prod}\AgdaSpace{}%
\AgdaBound{s}\AgdaSymbol{))}\AgdaSpace{}%
\AgdaSymbol{λ}\AgdaSpace{}%
\AgdaBound{i}\AgdaSpace{}%
\AgdaSymbol{→}\AgdaSpace{}%
\AgdaRecord{BFin}\AgdaSpace{}%
\AgdaSymbol{(}\AgdaNumber{1}\AgdaSpace{}%
\AgdaOperator{\AgdaPrimitive{+}}\AgdaSpace{}%
\AgdaFunction{blookup}\AgdaSpace{}%
\AgdaBound{v}\AgdaSpace{}%
\AgdaBound{i}\AgdaSymbol{)}\<%
\end{code}
where \texttt{blookup} has the type \texttt{Vec X n → BFin n → X} and it
selects an element from a vector at a given index.

Level-0 arrays can be cut only in a single way, producing two unit shapes.
Then the array of type \texttt{Ar 0 X tt} can be straight-forwardly nested
into the array of type \texttt{Ar 0 (Ar 0 X tt) tt)}.

Level-1 arrays can be nested in two different ways: with a singleton shape
on the outside or on the inside.  That is, $[1,2,3]$ can be turned into a
nested vector as $[[1,2,3]]$ or as $[[1],[2],[3]]$.

For arrays of levels greater than one we first pick an index into the shape of
the shape (vector \texttt{v}), and then we pick a number that is less or
equal than the element of \texttt{v} at that index.  Consider an
example.  For a level-2 array of shape \texttt{(tt , 3 ∷ []) , m ∷ n ∷ k ∷ []}
we first have to pick an element from the one-element vector, and then we
pick a number that is less or equal than 3.  Assume, we picked 1, in this
case all the elements of \texttt{m ∷ n ∷ k ∷ []} with index that is smaller to 1 will form
a left shape and the rest of the index will go into the right shape, resulting
in shapes \texttt{(tt , 1 ∷ []) , m ∷ []} and \texttt{(tt , 2 ∷ []) , n ∷ k ∷ []}.
Note that if we pick 0, we end-up with shapes \texttt{(tt , 0 ∷ []) , []}
and \texttt{(tt , 3 ∷ []) , m ∷ n ∷ k ∷ []}.  The left shape in this case
is a singleton which means that we will get a valid nesting.

We define a function that ``cuts'' a shape into two given the shape and
the argument of \texttt{RankedT} type.
\begin{code}%
\>[0]\AgdaPostulate{ranked-cut}\AgdaSpace{}%
\AgdaSymbol{:}\AgdaSpace{}%
\AgdaSymbol{∀}\AgdaSpace{}%
\AgdaSymbol{\{}\AgdaBound{l}\AgdaSpace{}%
\AgdaSymbol{:}\AgdaSpace{}%
\AgdaDatatype{ℕ}\AgdaSymbol{\}}\AgdaSpace{}%
\AgdaSymbol{→}\AgdaSpace{}%
\AgdaSymbol{(}\AgdaBound{s}\AgdaSpace{}%
\AgdaSymbol{:}\AgdaSpace{}%
\AgdaFunction{ShType}\AgdaSpace{}%
\AgdaBound{l}\AgdaSymbol{)}\AgdaSpace{}%
\AgdaSymbol{→}\AgdaSpace{}%
\AgdaFunction{RankedT}\AgdaSpace{}%
\AgdaBound{s}\AgdaSpace{}%
\AgdaSymbol{→}\AgdaSpace{}%
\AgdaFunction{ShType}\AgdaSpace{}%
\AgdaBound{l}\AgdaSpace{}%
\AgdaOperator{\AgdaFunction{×}}\AgdaSpace{}%
\AgdaFunction{ShType}\AgdaSpace{}%
\AgdaBound{l}\<%
\end{code}
And we use this function to define the nesting operation.
\begin{code}%
\>[0]\AgdaPostulate{nest}\AgdaSpace{}%
\AgdaSymbol{:}%
\>[576I]\AgdaSymbol{∀}\AgdaSpace{}%
\AgdaSymbol{\{}\AgdaBound{l}\AgdaSpace{}%
\AgdaBound{X}\AgdaSpace{}%
\AgdaBound{s}\AgdaSymbol{\}}\AgdaSpace{}%
\AgdaSymbol{→}\AgdaSpace{}%
\AgdaDatatype{Ar}\AgdaSpace{}%
\AgdaBound{l}\AgdaSpace{}%
\AgdaBound{X}\AgdaSpace{}%
\AgdaBound{s}\AgdaSpace{}%
\AgdaSymbol{→}\AgdaSpace{}%
\AgdaSymbol{(}\AgdaBound{ri}\AgdaSpace{}%
\AgdaSymbol{:}\AgdaSpace{}%
\AgdaFunction{RankedT}\AgdaSpace{}%
\AgdaBound{s}\AgdaSymbol{)}\AgdaSpace{}%
\AgdaSymbol{→}\<%
\\
\>[.][@{}l@{}]\<[576I]%
\>[7]\AgdaKeyword{let}\AgdaSpace{}%
\AgdaBound{s₁}\AgdaSpace{}%
\AgdaOperator{\AgdaInductiveConstructor{,}}\AgdaSpace{}%
\AgdaBound{s₂}\AgdaSpace{}%
\AgdaSymbol{=}\AgdaSpace{}%
\AgdaPostulate{ranked-cut}\AgdaSpace{}%
\AgdaBound{s}\AgdaSpace{}%
\AgdaBound{ri}\AgdaSpace{}%
\AgdaKeyword{in}\<%
\\
\>[7]\AgdaDatatype{Ar}\AgdaSpace{}%
\AgdaBound{l}\AgdaSpace{}%
\AgdaSymbol{(}\AgdaDatatype{Ar}\AgdaSpace{}%
\AgdaBound{l}\AgdaSpace{}%
\AgdaBound{X}\AgdaSpace{}%
\AgdaBound{s₂}\AgdaSymbol{)}\AgdaSpace{}%
\AgdaBound{s₁}\<%
\end{code}
The implementation details can be found in~\cite{github}.  Unfortunately
things get rather non-trivial rather quickly, as we have to implicitly prove
that the shapes obtained as a result of \texttt{ranked-cut} can be merged
together into the same shape.

Finally, consider an index-free formulation of the average pooling
using level-3 arrays.
\begin{code}%
\>[0]\AgdaPostulate{s≡s'}\AgdaSpace{}%
\AgdaSymbol{:}\AgdaSpace{}%
\AgdaSymbol{∀}\AgdaSpace{}%
\AgdaBound{m}\AgdaSpace{}%
\AgdaBound{n}\AgdaSpace{}%
\AgdaSymbol{→}%
\>[16]\AgdaBound{m}\AgdaSpace{}%
\AgdaOperator{\AgdaPrimitive{*}}\AgdaSpace{}%
\AgdaNumber{2}\AgdaSpace{}%
\AgdaOperator{\AgdaPrimitive{*}}\AgdaSpace{}%
\AgdaSymbol{(}\AgdaBound{n}\AgdaSpace{}%
\AgdaOperator{\AgdaPrimitive{*}}\AgdaSpace{}%
\AgdaNumber{2}\AgdaSpace{}%
\AgdaOperator{\AgdaPrimitive{*}}\AgdaSpace{}%
\AgdaNumber{1}\AgdaSymbol{)}\AgdaSpace{}%
\AgdaOperator{\AgdaDatatype{≡}}\AgdaSpace{}%
\AgdaBound{m}\AgdaSpace{}%
\AgdaOperator{\AgdaPrimitive{*}}\AgdaSpace{}%
\AgdaSymbol{(}\AgdaBound{n}\AgdaSpace{}%
\AgdaOperator{\AgdaPrimitive{*}}\AgdaSpace{}%
\AgdaNumber{2}\AgdaSpace{}%
\AgdaOperator{\AgdaPrimitive{+}}\AgdaSpace{}%
\AgdaSymbol{(}\AgdaBound{n}\AgdaSpace{}%
\AgdaOperator{\AgdaPrimitive{*}}\AgdaSpace{}%
\AgdaNumber{2}\AgdaSpace{}%
\AgdaOperator{\AgdaPrimitive{+}}\AgdaSpace{}%
\AgdaNumber{0}\AgdaSymbol{))}\<%
\\
\\[\AgdaEmptyExtraSkip]%
\>[0]\AgdaFunction{map}\AgdaSpace{}%
\AgdaSymbol{:}\AgdaSpace{}%
\AgdaSymbol{∀}\AgdaSpace{}%
\AgdaSymbol{\{}\AgdaBound{X}\AgdaSpace{}%
\AgdaBound{Y}\AgdaSpace{}%
\AgdaBound{l}\AgdaSpace{}%
\AgdaBound{s}\AgdaSymbol{\}}\AgdaSpace{}%
\AgdaSymbol{→}\AgdaSpace{}%
\AgdaSymbol{(}\AgdaBound{X}\AgdaSpace{}%
\AgdaSymbol{→}\AgdaSpace{}%
\AgdaBound{Y}\AgdaSymbol{)}\AgdaSpace{}%
\AgdaSymbol{→}\AgdaSpace{}%
\AgdaDatatype{Ar}\AgdaSpace{}%
\AgdaBound{l}\AgdaSpace{}%
\AgdaBound{X}\AgdaSpace{}%
\AgdaBound{s}%
\>[40]\AgdaSymbol{→}\AgdaSpace{}%
\AgdaDatatype{Ar}\AgdaSpace{}%
\AgdaBound{l}\AgdaSpace{}%
\AgdaBound{Y}\AgdaSpace{}%
\AgdaBound{s}\<%
\\
\>[0]\AgdaFunction{map}\AgdaSpace{}%
\AgdaBound{f}\AgdaSpace{}%
\AgdaSymbol{(}\AgdaInductiveConstructor{imap}\AgdaSpace{}%
\AgdaBound{a}\AgdaSymbol{)}\AgdaSpace{}%
\AgdaSymbol{=}\AgdaSpace{}%
\AgdaInductiveConstructor{imap}\AgdaSpace{}%
\AgdaSymbol{λ}\AgdaSpace{}%
\AgdaBound{iv}\AgdaSpace{}%
\AgdaSymbol{→}\AgdaSpace{}%
\AgdaBound{f}\AgdaSpace{}%
\AgdaOperator{\AgdaFunction{\$}}\AgdaSpace{}%
\AgdaBound{a}\AgdaSpace{}%
\AgdaBound{iv}\<%
\\
\\[\AgdaEmptyExtraSkip]%
\>[0]\AgdaFunction{avgp}%
\>[661I]\AgdaSymbol{:}\AgdaSpace{}%
\AgdaSymbol{∀}\AgdaSpace{}%
\AgdaSymbol{\{}\AgdaBound{m}\AgdaSpace{}%
\AgdaBound{n}\AgdaSymbol{\}}\<%
\\
\>[.][@{}l@{}]\<[661I]%
\>[5]\AgdaSymbol{→}\AgdaSpace{}%
\AgdaDatatype{Ar}\AgdaSpace{}%
\AgdaNumber{2}\AgdaSpace{}%
\AgdaDatatype{ℕ}\AgdaSpace{}%
\AgdaSymbol{((}\AgdaInductiveConstructor{tt}\AgdaSpace{}%
\AgdaOperator{\AgdaInductiveConstructor{,}}\AgdaSpace{}%
\AgdaNumber{2}\AgdaSpace{}%
\AgdaOperator{\AgdaInductiveConstructor{∷}}\AgdaSpace{}%
\AgdaInductiveConstructor{[]}\AgdaSymbol{)}\AgdaSpace{}%
\AgdaOperator{\AgdaInductiveConstructor{,}}\AgdaSpace{}%
\AgdaBound{m}\AgdaSpace{}%
\AgdaOperator{\AgdaPrimitive{*}}\AgdaSpace{}%
\AgdaNumber{2}\AgdaSpace{}%
\AgdaOperator{\AgdaInductiveConstructor{∷}}\AgdaSpace{}%
\AgdaBound{n}\AgdaSpace{}%
\AgdaOperator{\AgdaPrimitive{*}}\AgdaSpace{}%
\AgdaNumber{2}\AgdaSpace{}%
\AgdaOperator{\AgdaInductiveConstructor{∷}}\AgdaSpace{}%
\AgdaInductiveConstructor{[]}\AgdaSymbol{)}\<%
\\
\>[5]\AgdaSymbol{→}\AgdaSpace{}%
\AgdaDatatype{Ar}\AgdaSpace{}%
\AgdaNumber{2}\AgdaSpace{}%
\AgdaDatatype{ℕ}\AgdaSpace{}%
\AgdaSymbol{((}\AgdaInductiveConstructor{tt}\AgdaSpace{}%
\AgdaOperator{\AgdaInductiveConstructor{,}}\AgdaSpace{}%
\AgdaNumber{2}\AgdaSpace{}%
\AgdaOperator{\AgdaInductiveConstructor{∷}}\AgdaSpace{}%
\AgdaInductiveConstructor{[]}\AgdaSymbol{)}\AgdaSpace{}%
\AgdaOperator{\AgdaInductiveConstructor{,}}\AgdaSpace{}%
\AgdaBound{m}\AgdaSpace{}%
\AgdaOperator{\AgdaInductiveConstructor{∷}}\AgdaSpace{}%
\AgdaBound{n}\AgdaSpace{}%
\AgdaOperator{\AgdaInductiveConstructor{∷}}\AgdaSpace{}%
\AgdaInductiveConstructor{[]}\AgdaSymbol{)}\<%
\\
\>[0]\AgdaFunction{avgp}\AgdaSpace{}%
\AgdaSymbol{\{}\AgdaBound{m}\AgdaSymbol{\}\{}\AgdaBound{n}\AgdaSymbol{\}}\AgdaSpace{}%
\AgdaBound{a}\AgdaSpace{}%
\AgdaSymbol{=}\AgdaSpace{}%
\AgdaKeyword{let}\<%
\\
\>[0][@{}l@{\AgdaIndent{0}}]%
\>[2]\AgdaBound{s₁}\AgdaSpace{}%
\AgdaSymbol{:}\AgdaSpace{}%
\AgdaFunction{ShType}\AgdaSpace{}%
\AgdaNumber{3}\<%
\\
\>[2]\AgdaBound{s₁}%
\>[704I]\AgdaSymbol{=}\AgdaSpace{}%
\AgdaSymbol{((\AgdaUnderscore{}}\AgdaSpace{}%
\AgdaOperator{\AgdaInductiveConstructor{,}}\AgdaSpace{}%
\AgdaNumber{2}\AgdaSpace{}%
\AgdaOperator{\AgdaInductiveConstructor{∷}}\AgdaSpace{}%
\AgdaInductiveConstructor{[]}\AgdaSymbol{)}\AgdaSpace{}%
\AgdaOperator{\AgdaInductiveConstructor{,}}\AgdaSpace{}%
\AgdaNumber{2}\AgdaSpace{}%
\AgdaOperator{\AgdaInductiveConstructor{∷}}\AgdaSpace{}%
\AgdaNumber{2}\AgdaSpace{}%
\AgdaOperator{\AgdaInductiveConstructor{∷}}\AgdaSpace{}%
\AgdaInductiveConstructor{[]}\AgdaSymbol{)}\<%
\\
\>[704I][@{}l@{\AgdaIndent{0}}]%
\>[6]\AgdaOperator{\AgdaInductiveConstructor{,}}%
\>[9]\AgdaBound{m}\AgdaSpace{}%
\AgdaOperator{\AgdaInductiveConstructor{∷}}\AgdaSpace{}%
\AgdaNumber{2}\AgdaSpace{}%
\AgdaOperator{\AgdaInductiveConstructor{∷}}\AgdaSpace{}%
\AgdaBound{n}\AgdaSpace{}%
\AgdaOperator{\AgdaInductiveConstructor{∷}}\AgdaSpace{}%
\AgdaNumber{2}\AgdaSpace{}%
\AgdaOperator{\AgdaInductiveConstructor{∷}}\AgdaSpace{}%
\AgdaInductiveConstructor{[]}\<%
\\
\>[2]\AgdaBound{a₁}\AgdaSpace{}%
\AgdaSymbol{=}\AgdaSpace{}%
\AgdaPostulate{reshape}\AgdaSpace{}%
\AgdaSymbol{\{}\AgdaArgument{s₁}\AgdaSpace{}%
\AgdaSymbol{=}\AgdaSpace{}%
\AgdaBound{s₁}\AgdaSymbol{\}}\AgdaSpace{}%
\AgdaBound{a}\AgdaSpace{}%
\AgdaSymbol{(}\AgdaPostulate{s≡s'}\AgdaSpace{}%
\AgdaBound{m}\AgdaSpace{}%
\AgdaBound{n}\AgdaSymbol{)}\<%
\\
\>[2]\AgdaBound{aₙ}\AgdaSpace{}%
\AgdaSymbol{=}\AgdaSpace{}%
\AgdaPostulate{nest}\AgdaSpace{}%
\AgdaBound{a₁}\AgdaSpace{}%
\AgdaSymbol{((}\AgdaNumber{1}\AgdaSpace{}%
\AgdaOperator{\AgdaInductiveConstructor{bounded}}\AgdaSpace{}%
\AgdaFunction{auto≥}\AgdaSymbol{)}\AgdaSpace{}%
\AgdaOperator{\AgdaInductiveConstructor{,}}\AgdaSpace{}%
\AgdaSymbol{(}\AgdaNumber{1}\AgdaSpace{}%
\AgdaOperator{\AgdaInductiveConstructor{bounded}}\AgdaSpace{}%
\AgdaFunction{auto≥}\AgdaSymbol{))}\<%
\\
\>[2]\AgdaBound{r₃}\AgdaSpace{}%
\AgdaSymbol{=}\AgdaSpace{}%
\AgdaFunction{map}\AgdaSpace{}%
\AgdaSymbol{((}\AgdaOperator{\AgdaFunction{\AgdaUnderscore{}/}}\AgdaSpace{}%
\AgdaNumber{4}\AgdaSymbol{)}\AgdaSpace{}%
\AgdaOperator{\AgdaFunction{∘}}\AgdaSpace{}%
\AgdaFunction{sum}\AgdaSymbol{)}\AgdaSpace{}%
\AgdaBound{aₙ}\<%
\\
\>[2]\AgdaKeyword{in}\AgdaSpace{}%
\AgdaPostulate{reshape}\AgdaSpace{}%
\AgdaBound{r₃}\AgdaSpace{}%
\AgdaInductiveConstructor{refl}\<%
\end{code}
First, we reshape the input array of shape \texttt{m*2 ∷ n*2 ∷ []} into the level-3
array of shape
\(    s₁ =
      \left(
      \begin{matrix}
        m & 2 \\
        n & 2
      \end{matrix}
      \right)
\).  Then we cut $s₁$ using \texttt{ranked-cut} with the argument (1 , 1)
of type \texttt{RankedT}.  The first `1' says that from axes \texttt{2 ∷ 2 ∷ []}
(the shape of $s₁$) we pick the second.  The second `1' says that
all the $s₁$ elements which have index where the second component is smaller
than one will form the left shape of the \texttt{ranked-cut}.  In other words
we are saying that we are cutting $s₁$ vertically, taking the first column
as the left shape.  Then we apply the average function to
\(
      \left(
      \begin{matrix}
        2 \\
        2
      \end{matrix}
      \right)
\)
subarrays.  The \texttt{auto≥} is an instance function that automatically
proves trivial inequalities on natural numbers.  The \texttt{map} is defined
right before the \texttt{avgp}.  Note that this version of the map does not
prescribe the order in which array elements are traversed. 
We say only that all the elements are modified with some function \texttt{f} of the
appropriate type.  In the last statement, the equality between products of
\(
      \left(
      \begin{matrix}
        2 \\
        2
      \end{matrix}
      \right)
\)
level-3 shape and $2\times 2$ level-2 shape is obvious to Agda.

\paragraph{Discussion}
The definition of the \texttt{nest} operator uses some quite heavy machinery,
even though we eliminated direct index manipulations in the average function.
One can ask whether this is justified.  Surely, for the purposes
of a single function it may be not.  However, the pattern exemplified above,
in which we
reshape an array in a such a way that inner dimensions contain the
elements of interest and then we map a function over all these elements
is extremely common in array programming.  In an array library
or a DSL embedded into Agda-like framework, the \emph{ability to define}
the extended ranked operator is very valuable --- we only have to define
it once.

Finally, while defining \texttt{nest} we had to first prove the fact that
index to offset and offset to index functions are inverses of each other.
The core of this idea is captured in the theorem \texttt{io-oi} 
in~\cite{github}.

\section{Conclusions}

We have presented a novel data structure that generalises multi-dimensional
arrays.  The key to our construction is a strong symmetry or analogy
between the type that describes the shape of the data structure and the data structure
itself.  Such a symmetry gives rise to the hierarchy of types --- in our case,
an array of natural numbers can be used as a shape descriptor of the next array
type in the hierarchy.  We start with unit shapes and corresponding one-element
level-0 arrays.  After that, level-1 arrays have a shape that is described by
a single natural number and the array itself has $n$ elements.  At level 2 we
the shape is given by an $n$-element array of natural numbers, and so on.

This hierarchy appeared naturally after we encoded multi-dimensional arrays
using containers.  We discovered a new container operation, and the iteration
of this operation led to the array type hierarchy.

While only the first three levels of the hierarchy have been used in practice so
far, we have demonstrated that higher levels are also of practical useful.  They
naturally fit the tradition of rank-polymorphic aggregated operations in
the style of APL, suggesting that array operations can be expressed in
an index-free combinator style.  The availability of higher-level arrays
makes it possible to enrich the functionality of existing combinators, as
we have demonstrated at the example of the ranked operator.

In order to make this idea precise, we encoded the proposed array types
and operations in Agda and observed a number of standard array properties.

This work opens up a number of interesting research directions.  For example:
using other containers to form similar type hierarchies.  It would be interesting
to explore alternative ways to work around the container extensionality problem,
so that less encoding machinery needs to be exposed.

\section*{Acknowledgements}
I am very grateful to Peter Hancock.  He was the first one to suggest to treat
arrays as containers, and came up with the initial version of the $\diamond$
operation.  Two days of intensive discussions with Sven-Bodo Scholz on the
meaning of arrays with levels were of a great help as well.
\bibliographystyle{eptcs}
\bibliography{paper}
\end{document}